\documentclass[12pt]{article}
\usepackage{graphicx}
\usepackage{longtable}
\makeatletter
 
  \@addtoreset{equation}{section}
\makeatother
\def\beqar {\begin{eqnarray}}
\def\eeqar {\end{eqnarray}}
\def\beq {\begin{equation}}
\def\eeq {\end{equation}}
\def\A{{\cal A}}

\def\C{{\cal C}}
\def\F{{\cal F}}

\def\S{{\cal S}}

\def\P{{\cal P}}

\def\N{{\cal N}}
\def\M{{\cal M}}

\def\al{\alpha}
\def\bt{\beta}
\def\del{\delta}

\def\ga{\gamma}

\def\ep{\epsilon}
\def\la{\lambda}

\def\om{\omega}

\def\th{\theta}

\def\si{\sigma}

\def\Ad{{\dot A}}
\def\Bd{{\dot B}}

\def\bu{{\bar u}}

\def\hf{\frac{1}{2}}

\def\<{\langle}\def\bra{\langle}
\def\>{\rangle}\def\ket{\rangle}

\def\Tr{{\rm Tr}}

\def\Path{{\rm P}}

\def\cp{{\bf CP}}

\begin{document}

\begin{titlepage}
\null\vspace{-62pt} \pagestyle{empty}
\begin{center}
%\rightline{} \rightline{CCNY-HEP-/05}
\vspace{1.0truein}

{\Large\bf Holonomies of gauge fields in twistor space 3: \\
\vspace{.35cm}
\hspace{-.3cm}
gravity as a square of ${\cal N}=4$ theory} \\

%%%%%%%%%%%%%%%%%%%%%%%%%%%%%%%%%%%%%%%%%%%%%%%%%%%%%%%%%%
\vspace{1.0in} {\sc Yasuhiro Abe} \\
\vskip .12in {\it Cereja Technology Co., Ltd.\\
1-13-14 Mukai-Bldg. 3F, Sekiguchi \\
Bunkyo-ku, Tokyo 112-0014, Japan } \\
\vskip .07in {\tt abe@cereja.co.jp}\\
\vspace{1.3in}
%%%%%%%%%%%%%%%%%%%%%%%%%%%%%%%%%%%%%%%%%%%%%%%%%%%%%%%%%%%%
\centerline{\large\bf Abstract}
\end{center}
In a recent paper, we show that an S-matrix functional for graviton amplitudes
can be described by an ${\N = 8}$ supersymmetric gravitational holonomy operator
in twistor space.
In this paper, we obtain an alternative expression for the gravitational
holonomy operator such that it can be interpreted as a square of
an ${\N = 4}$ holonomy operator for frame fields, by taking a sum
of certain shuffles over ordered indices.
The new expression leads to amplitudes of not only spin-$2$ gravitons
but also spin-$0$ massless particles.
We discuss that the squared model is favored as a theory of quantum gravity.

\end{titlepage}
%%%%%%%%%%%%%%%%%%%%%%%%%%%%%%%%%%%%%%%%%%%%%%%%%%%%%%%%%%%%%
\pagestyle{plain} \setcounter{page}{2} %\baselineskip =14pt

%%%%%%%%%%%%%%%%%%%%%%%%%%%%%%%%%%%%%%%%%%%%%%%%%%%%%%%%%%%%%%%%%%
\section{Introduction}

In recent years there has been much attention to the relation
between $\N = 4$ super Yang-Mills theory and $\N = 8$ supergravity.
An explicit relation between the two theories at the level
of classical scattering amplitudes is first obtained
in \cite{Berends:1988zp} by taking the field-theory limit of
the so-called Kawai-Lewellen-Tye (KLT) relation between
tree-level amplitudes of open and closed string theories \cite{Kawai:1985xq}.
Roughly speaking, this relation gives an expression of graviton amplitudes
in terms of a square of gluon counterparts, with certain multiplicity factors.
Field theoretic construction of a gravitational theory by use of this relation
has been studied in earlier works of Bern and others \cite{Bern:1998sv,Bern:2002kj}.
What is suggested in Bern's approach is that one may reduce the ultraviolet
behavior of $\N = 8$ supergravity to that of $\N = 4$ super Yang-Mills theory
if one utilizes structural similarities between the two theories.

It is not until the work of Witten \cite{Witten:2003nn}, which
generalizes Nair's observation on the so-called maximally helicity
violating (MHV) amplitudes of gluons in a twistor-space framework \cite{Nair:1988bq},
that many researchers start realizing that Bern's approach is
in fact very promising in showing the ultraviolet finiteness of $\N = 8$ supergravity
as a theory of quantum gravity.
This is partly because recent developments in the helicity-based calculation
of gluon amplitudes show that the amplitudes can significantly be
simplified, even at loop levels, by use of the MHV amplitudes (or vertices).
It is therefore natural to apply these developments to gravitational
theories using the above-mentioned relation between gauge theory and gravity.
In fact, there are a plentiful number of papers on this specific subject.
For some earlier works, see for example
\cite{BjerrumBohr:2004wh}-\cite{ArkaniHamed:2008yf}.
For very recent papers, see also \cite{BjerrumBohr:2010zp}-\cite{Bern:2010fy}.

In the present paper, following these lines of developments,
we investigate the ``squared'' relation between gauge theory and gravity
in a recently proposed holonomy formalism \cite{Abe:2009kn,Abe:2009kq}.
In \cite{Abe:2009kq} we construct a gravitational holonomy operator in twistor space,
interpreting gravity as a gauge theory with nontrivial
Chan-Paton factors.
We then show that an S-matrix functional for
graviton amplitudes can be expressed in terms of a supersymmetric
version of the holonomy operator.
Motivated by Bern's approach, in this paper we shall change our interpretation
of gravity to obtain an alternative expression for the gravitational
holonomy operator such that we can easily understand it
as a square of a gauge-theory holonomy operator with $\N = 4$ extended supersymmetry.
We shall also check that the alternative expression does reproduce
the correct graviton amplitudes.

This paper is organized as follows. In the next section, we review
the construction of a gravitational holonomy operator, following \cite{Abe:2009kq},
and present its explicit definition.
In section 3, we treat the summations that appear in the gravitational holonomy operator
in a different manner so that it can easily be regarded as
a square of a gauge-theory holonomy operator.
In section 4, we consider supersymmetrization of the holonomy operators
and confirm that the new expression also correctly leads to graviton amplitudes.
Lastly, we shall present some concluding remarks.

%%%%%%%%%%%%%%%%%%%%%%%%%%%%%%%%%%%%%%%%%%%%%%%%%%%%%%%%%%%
\section{Review of a gravitational holonomy operator}

\noindent
\underline{Definition}

In this section we review the construction of a gravitational holonomy operator
which is proposed in a recent paper \cite{Abe:2009kq}.
The gravitational holonomy operator is defined by
\beq
    \Theta_{R, \ga}^{(H)} (u, \bu) = \Tr_{R, \ga} \, \Path \exp \left[
    \sum_{m \ge 3} \oint_{\ga} \underbrace{H \wedge H \wedge \cdots \wedge H}_{m}
    \right]
\label{2-1}
\eeq
where $H$ is called a comprehensive graviton field and is defined by the
following set of equations.
\beqar
    H &=& \sqrt{8 \pi G_N}
    \sum_{1 \le i < j \le m} \,
    \sum_{\si \in \S_{r-1}}
    \sum_{\tau \in \S_{m-r-2}}
    \left(
    \sum_{h_{i \mu_i}} g_{i}^{(h_{i \mu_i})} \otimes g_{j}^{(00)}
    \right)
    \, \om_{ij} \, \om_{\la_{i} \la_{j}}
    \label{2-2} \\
    \mu_i & = &  \left\{
        \begin{array}{ll}
          \si_{i} & \mbox{for $i = 1, 2, \cdots , r$}\\
          \tau_{i}  & \mbox{for $i = r+1, r+2, \cdots, m-1$} \\
          m  & \mbox{for $i = m$} \\
        \end{array} \right.
    \label{2-3}\\
    \la_i & = &  \left\{
        \begin{array}{ll}
          \si_{i+1} & \mbox{for $i = 1, 2, \cdots , r$}\\
          \tau_{i-1}  & \mbox{for $i = r+1, r+2, \cdots, m-1$} \\
          m  & \mbox{for $i = m$} \\
        \end{array} \right.
    \label{2-4} \\
    g_{i}^{(h_{i \mu_i})} & = &   T^{\mu_i} \, g_{i \mu_i}^{(h_{i \mu_i})}
    \, = \, T^{\mu_i} \, e_{i}^{(h_{i}) a} \, e_{\mu_i}^{(h_{\mu_i}) a}
    \label{2-5} \\
    g_{j}^{(00)} & = &  ( {\bf 1} )^{\mu_j} \, e_{j}^{(0)} \, e_{\mu_j}^{(0)}
    \label{2-6}
\eeqar

In the rest of this section,
we shall explain the notations of the above expressions one by one.
The reader may find the following description lengthy but what we
shall do is nothing but to present the definition of
quantum gravity in the holonomy formalism.
Thus we find it important to review the definition in a consistent
manner. We try to make the discussion as much concise as possible;
for details of the definition, the reader may refer to \cite{Abe:2009kq}.

\vskip 5mm
\noindent
\underline{Coupling constant, numbering indices and braid diagrams}

First of all, $G_{N}$ denotes the Newton constant.
In the natural unit ($ c = \hbar = 1$), this is equivalent to
the inverse square of the Planck mass $M_{Pl}$:
\beq
    G_{N} = \frac{1}{M_{Pl}^{2}} = 6.7088 \times 10^{-39} \,
    \left[ \frac{1}{\mbox{GeV}} \right]^2 \, .
    \label{2-7}
\eeq

\begin{figure} [htbp]
\begin{center}
\includegraphics[width=140mm]{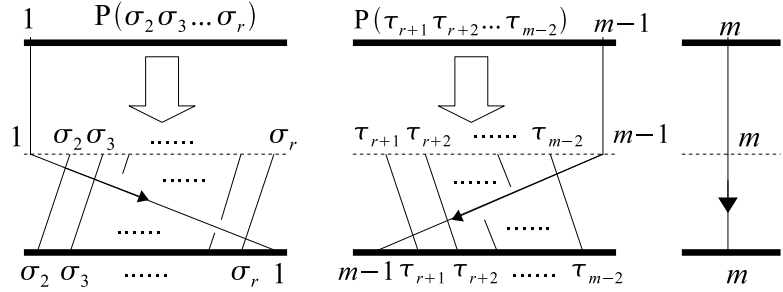}
\caption{Braid diagrams corresponding to the permutations of $\si$'s and $\tau$'s
as well as the identity permutation of the index $m$
--- the symbol $\Path$ denotes an ascending ordering of the arguments.
When two lines are crossing each other, we consider that a line with
an arrow is closer to us, crossing over the other line without an arrow.}
\label{fig02}
\end{center}
\end{figure}

The numbering indices $i$, $j$ take values of $1, 2 ,\cdots , m$.
We split these into $\{ 2, 3, \cdots , r \}$ and
$\{ r+ 1 , r+ 2 , \cdots , m -1 \}$ ($2 \le r \le m-2$)
and consider transpositions (or permutations) of the two distinct sets of indices.
The transpositions are labeled by
\beq
    \si=\left(%
    \begin{array}{c}
      2 \cdots r \\
      \si_2 \cdots \si_r \\
    \end{array}%
    \right) \, , ~~~
    \tau=\left(%
    \begin{array}{c}
      r+1 \cdots m-2 \\
      \tau_{r+1} \cdots \tau_{m-2} \\
    \end{array}%
    \right) \, .
    \label{2-8}
\eeq
The sum of the transpositions $\si$ can be denoted as
a sum over $\si \in \S_{r-1}$ where $\S_{r-1}$ represents
the rank-$(r-1)$ symmetric group.
Similarly the sum of the transpositions $\tau$ can be denoted
by a sum over $\tau \in \S_{m-r-2}$.
We fix the rest of the numbering indices, $1$, $m-1$ and $m$,
out of the permutations.
For convenience, we denote this fact by
\beq
    \si_1 = \si_{r+1} = 1 \, , ~~~ \tau_{m-1} = \tau_{r} = m-1 \,  .
    \label{2-9}
\eeq
The above permutations of the indices are schematically shown in
Figure \ref{fig02} where we draw braid diagrams for $\si$'s and $\tau$'s separately.
In the figure, the elements of $\si$'s and $\tau$'s
are chosen at random, while the symbol $\Path$ denotes an
ascending ordering of the elements.
Structure of each braid diagram depends on a specific
choice of the permutation; the structure is shown by
a thick down-arrow in Figure \ref{fig02}.

%%%%%%%%%%%%%%%%%%%%%%%%% figure %%%%%%%%%%%%%%%%%%%%%%%%%
%%%%%%%%%%%%%%%%%%%%%%%%% figure %%%%%%%%%%%%%%%%%%%%%%%%%

The index $\mu_i$ in (\ref{2-3}) is a composite numbering index
in a sense that it covers both $\si$'s and $\tau$'s.
Similarly the index $\la_i$ in (\ref{2-4}) is a composite numbering index;
the difference from $\mu_i$ is that the indices of $\si$'s
are positively shifted by one while those of $\tau$'s are negatively shifted by
one.
For the full definition of $H$, we need to define
\beq
    \la_{r} = 1 \, , ~~ \la_{r+1} = m-1 \, , ~~ \la_{m+1} = \si_2 \, .
    \label{2-10}
\eeq
The first two relations are in accord with (\ref{2-9}).
Information of $\la_{m+1}$ is necessary in defining the gravitational
holonomy operator (\ref{2-1}).

\noindent
\underline{Chan-Paton factors of gravitons and frame fields}

Now we explain the meaning of the graviton operator
$g_{i}^{(h_{i \mu_i})}$ in (\ref{2-5}).
A graviton labeled by a particular
numbering index corresponds to a particular strand in the braid diagrams
in Figure \ref{fig02}.
$h_{i \mu_i}$ represents the helicity of the $i$-th graviton, taking
a value of $h_{i \mu_i} \equiv h_{i} h_{\mu_i} = (++, +-, -+, --)$ where
$h_{i}$ and $h_{\mu_i}$, both taking a value of $\pm$, denote ``helicities'' of
the frame fields $e_{i}^{(h_i )a}$ and $e_{\mu_i}^{( h_{\mu_i} )a}$,
respectively. Here $a$ $(=0,1,2,3)$ represents a tangent-space index.

The Chan-Paton factor $T^{\mu_i}$ of the graviton $g_{i}^{(h_{i \mu_i})}$
is determined by a specific choice of the permutations, $\si$ and $\tau$.
Namely, the factor $T^{\mu_i}$ is in one-to-one correspondence with
$( \si , \tau )$ and can be defined as
\beqar
      T^{\mu_i} & = &  \left\{
        \begin{array}{ll}
         T^{\si_i} & \mbox{for $i = 1, 2, \cdots , r$}\\
         T^{\tau_{i}}  & \mbox{for $i = r+1, r+2, \cdots, m-1$} \\
         T^{m}  & \mbox{for $i = m$} \\
       \end{array} \right.
      \label{2-11}\\
      T^{\si_{i}} &=& \left\langle (p_1 + p_{\si_{i+1 < i}} + p_{\si_{i+2 < i}} +
      \cdots + p_{\si_{r < i}} ) \cdot p_{\si_i} \right\rangle
      \nonumber \\
      & =& \left\langle \left( p_1 + \sum_{k=i+1}^{r} p_{\si_{k < i}} \right)
      \cdot p_{\si_{i}} \right\rangle
      \label{2-12}\\
      T^{\tau_{i}} &=&
      \left\langle p_{\tau_i}  \cdot (p_{m-1} + p_{\tau_{i < r+1}} + p_{\tau_{i < r+2}} +
      \cdots + p_{\tau_{i < i -1 }} )  \right\rangle
      \nonumber \\
      &=& \left\langle p_{\tau_i} \cdot
      \left( p_{m-1} + \sum_{k = r+1}^{i-1} p_{\tau_{i < k}} \right)
      \right\rangle
      \label{2-13}\\
      T^{m}  &=& 1 ~ = ~ T^{1} ~=~ T^{m-1}
      \label{2-14}
\eeqar
where $\bra p_{i} \cdot p_{j} \ket$ represents a product of
four-momenta for the $i$-th and the $j$-th gravitons.
The meaning of the bracket will be clarified in a moment (see (\ref{2-24}) for the definition).
We fix $T^{1}$, $T^{m-1}$ and $T^{m}$ to the identity. This is related to
the fact that the spinor momenta $u$ preserves the $SL (2, {\bf C}) $ symmetry
which we discuss later.
In the above expressions, $p_{\si_{i<j}} $ and $p_{\tau_{i<j}}$ are defined as follows.
\beq
    p_{\si_{i<j}}  \, = \,   \left\{
        \begin{array}{ll}
          p_{\si_i} & \mbox{for $\si_i < \si_j$}\\
          0  & \mbox{otherwise}
        \end{array} \right.
    \, ~~~~
    p_{\tau_{i<j}}  \, = \,   \left\{
        \begin{array}{ll}
          p_{\tau_i} & \mbox{for $\tau_i < \tau_j$}\\
          0  & \mbox{otherwise}
        \end{array} \right.
    \label{2-15}
\eeq

The Chan-Paton factors of gravitons are
expressed in terms of the products of four-momenta,
with certain combinatoric structures.
This is natural if we notice that a graviton is composed of
two frame fields and that their Chan-Paton factors are
given by translational operators on the tangent spaces.
An explicit form of the frame-filed operator $e_{i}^{(\pm)}$
can be defined as
\beq
    e_{i}^{(\pm)} = e_{i}^{(\pm) a} (\sqrt{2} p_{i})^{a}
    = e_{i}^{(\pm) A \Ad} \, p_{i}^{A \Ad}
    \label{2-16}
\eeq
where we split the tangent-space index $a$ ($= 0,1,2,3$)
into the two-component indices $A$ and $\Ad$ both of which
take values of $(1,2)$.
As is seen in a moment,
the factor of $\sqrt{2}$ arises from the use of spinor momenta.
$p_{i}^{A \Ad}$ is a translational operator in the tangent space.
Since the tangent space is generally given by a copy of the
coordinate space, we can interpret $p_{i}^{A \Ad}$
as the four-momentum of the $i$-th graviton.

\noindent
\underline{Spinor momenta, twistor space and products of four-momenta}

Since $p_{i}^{A \Ad}$ satisfies the on-shell condition $p_{i}^{2} = 0$,
it can be written in terms of two-component spinor momenta $u_{i}^{A}$
and $\bu_{i}^{\Ad}$.
Explicitly, this can be written as
\beq
    p_{i}^{A \Ad} = (\si^{a})^{A \Ad} p_{i \, a}
    = u_{i}^{A} \bu_{i}^{\Ad }
    \label{2-17}
\eeq
where $\si^{a}=( {\bf 1} , {\vec \si} )$, with
${\vec \si}$ and ${\bf 1}$ denoting the ordinary $(2 \times 2)$ Pauli matrices and
the $(2 \times 2)$ identity matrix, respectively.
Explicit forms of the spinor momenta are then given by
\beq
    u^A = {1 \over \sqrt{p_0 - p_3}} \left(
        \begin{array}{c}
          {p_1 - i p_2} \\
          {p_0 - p_3} \\
        \end{array}
      \right) \, , ~~~
    \bu_\Ad  = {1 \over \sqrt{p_0 - p_3}}
    \left(
         \begin{array}{c}
           {p_1 + i p_2 } \\
           {p_0 - p_3} \\
         \end{array}
    \right)
    \label{2-18}
\eeq
where we omit the numbering index for simplicity.
Notice that we can take $\bu_\Ad$ as a
conjugate of $u^A$, {\it i.e.}, $\bu_\Ad = (u^A)^*$
by requiring that the four-momenta are real.

Lorentz transformations of $u^A$ are given by
\beq
    u^A \rightarrow (g u)^A
    \label{2-19}
\eeq
where $g \in SL(2, {\bf C})$
is a $(2 \times 2)$-matrix representation of $SL(2,{\bf C})$;
the complex conjugate of this relation leads to Lorentz transformations of $\bu_\Ad$.
Four-dimensional Lorentz transformations are realized by a
combination of these, that is, the four-dimensional Lorentz symmetry is
given by $SL(2,{\bf C}) \times SL(2,{\bf C})$.
Scalar products of $u^A$'s or $\bu_\Ad$'s, which are invariant under the
corresponding $SL(2,{\bf C})$, are expressed as
\beq
    u_i \cdot u_j \equiv (u_i u_j) =   \ep_{AB} u_{i}^{A}u_{j}^{B} \, , ~~~~~
    \bu_i \cdot \bu_j \equiv [\bu_i \bu_j]  = \ep^{\Ad \Bd} \bu_{i \, \Ad}
    \bu_{j \, \Bd}
    \label{2-20}
\eeq
where $\ep_{AB}$ is the rank-2 Levi-Civita tensor.
This can be used to raise or lower the indices, {\it e.g.}, $u_B = \ep_{AB}u^A$.
Notice that these products are zero when $i$ and $j$ are identical.

For a theory with conformal invariance, such as a theory of electromagnetism
or $\N = 4$ super Yang-Mills theory, we can impose scale invariance
on the spinor momentum, {\it i.e.},
\beq
    u^A \sim \la u^A \, , ~~~~~ \la \in {\bf C} - \{ 0 \}
    \label{2-21}
\eeq
where $\la$ is non-zero complex number.
With this identification, we can regard
the spinor momentum $u^A$ as a homogeneous coordinate
of the complex projective space $\cp^1$.
In the spinor-momenta formalism, we identify this $\cp^1$ as
an $S^2$ fiber of the twistor space $\cp^3$.
In this formulation, four-dimensional spacetime coordinates $x_{\Ad A}$
emerges form the twistor-space condition
\beq
    v_{\Ad} \, = \, x_{\Ad A} u^A
    \label{2-22}
\eeq
where $v_\Ad$ is another two-component complex spinor and the twistor space is defined by a
four-component spinor $Z_I =( u^A, v_\Ad)$ $(I = 1,2,3,4)$ that satisfies the scale invariance
\beq
    Z_I  \sim \la Z_I \, , ~~~~~ \la \in {\bf C} - \{ 0 \} \, .
    \label{2-23}
\eeq

In terms of the spinor momenta, products of four-momenta in the form of (\ref{2-17})
can be defined as
\beq
    p_{i}^{A \Ad} p_{j \, \Ad A} = (u_i u_j) [\bu_i \bu_j] = 2 p_{i}^{a} p_{j \, a}
    \equiv \bra p_i \cdot p_j \ket
    \label{2-24}
\eeq
where we use the expressions in (\ref{2-20}).
This shows an explicit meaning of the brackets that appear in (\ref{2-12}) and (\ref{2-13}).
It also explains the factor $\sqrt{2}$ in (\ref{2-16}).

\noindent
\underline{Frame-field holonomy: bialgebraic operator and logarithmic one-form}

As mentioned earlier, the sign $\pm$ in (\ref{2-16})
represents an analog of helicity for the frame field.
In fact, in our construction we consider the frame field as
a massless gluon, with its Chan-Paton factor specified by (\ref{2-16}).
This means that we can define a holonomy operator for the frame field:
\beqar
    \Theta_{R, \ga}^{(E)} (u) &=& \Tr_{R, \ga} \, \Path \exp \left[
    \sum_{m \ge 2} \oint_{\ga} \underbrace{E \wedge E \wedge \cdots \wedge E}_{m}
    \right]
    \label{2-25} \\
    E &=& \sum_{1 \le i < j \le n} E_{ij} \om_{ij}
    \label{2-26}\\
    E_{ij} &=& e_{i}^{(+)} \otimes e_{j}^{(0)}
    + e_{i}^{(-)} \otimes e_{j}^{(0)}
    \label{2-27}\\
    \om_{ij} &=& d \log (u_i u_j) = \frac{ d (u_i u_j) }{(u_i u_j) }
    \label{2-28}
\eeqar
where the operators $e_{i}^{(\pm)}$ and $e_{i}^{(0)}$
obey the $SL(2 , {\bf C})$ algebra. Explicitly this can be expressed as
\beq
    [ e_{i}^{(+)}, e_{j}^{(-)}] = 2 e_{i}^{(0)} \, \del_{ij}  \, , ~~~
    [ e_{i}^{(0)}, e_{j}^{(+)}] = e_{i}^{(+)} \, \del_{ij} \, , ~~~
    [ e_{i}^{(0)}, e_{j}^{(-)}] = - e_{i}^{(-)} \, \del_{ij}
    \label{2-29}
\eeq
where Kronecker's deltas show that
the non-zero commutators are obtained only for $i = j$.
The remaining of commutators, those expressed otherwise, all vanish.
As shown in (\ref{2-28}), $\om_{ij}$ is a logarithmic one-form in terms of the
Lorentz invariant product of the spinor momenta $u_i$ and $u_j$.
$\om_{ij}$ in (\ref{2-2}) is also defined by this logarithmic one-form.
The bialgebraic operator $E$ in (\ref{2-26}) is what we may call
the comprehensive frame field.

\noindent
\underline{Configuration space, ``path'' ordering and braid trace}

Physical variables of the comprehensive frame field $E$ is given by
$n$ spinor momenta. Since these are symmetric to each other, a physical
configuration space of $E$ and hence that of $\Theta_{R, \ga}^{(E)} (u)$
can be defined by $\C = {\bf C}^{n} / \S_{n}$ where
$\S_n$ denotes the rank-$n$ symmetric group.
The symbol $\ga$ in (\ref{2-25}) represents a closed path on $\C$
along which the integral is evaluated.
On the other hand, the symbol $R$ in (\ref{2-25}) denotes
the representation of the algebra of the Chan-Paton factor.

The symbol $\Path$ in (\ref{2-25}) denotes an ordering of the numbering indices.
The meaning of the action of $\Path$ on the exponent
of (\ref{2-25}) can explicitly be written as
\beqar
    \nonumber
    \Path \sum_{m \ge 2}  \oint_{\ga} \underbrace{E \wedge E \wedge \cdots \wedge E}_{m}
    &=& \sum_{m \ge 2} \oint_{\ga}  E_{1 2} E_{2 3} \cdots E_{m 1}
    \, \om_{12} \wedge \om_{23} \wedge \cdots \wedge \om_{m 1} \\
    \nonumber
    &=& \sum_{m \ge 2}  \frac{1}{2^{m+1}} \sum_{(h_1, h_2, \cdots , h_m)}
    (-1)^{h_1 + h_2 + \cdots + h_m} \\
    && ~~~ \times \,
    e_{1}^{(h_1)} \otimes e_{2}^{(h_2)} \otimes \cdots \otimes e_{m}^{(h_m)}
    \, \oint_{\ga} \om_{12} \wedge \cdots \wedge \om_{m1}
    \label{2-30}
\eeqar
where $h_{i} = \pm = \pm 1$ ($i=1,2,\cdots, m$)
denotes the ``helicity'' of the $i$-th frame field.
In obtaining the above expression, we use an ordinary definition of commutators for
bialgebraic operators. For example, using the commutation relations (\ref{2-29}),
we can calculate $[ E_{12} , E_{23} ]$ as
\beqar
    [E_{12}, E_{23}]
    &=& e_{1}^{(+)} \otimes e_{2}^{(+)} \otimes e_{3}^{(0)}
    - e_{1}^{(+)} \otimes e_{2}^{(-)} \otimes e_{3}^{(0)}
    \nonumber\\
    && \!\!\! + \, e_{1}^{(-)} \otimes e_{2}^{(+)} \otimes e_{3}^{(0)}
    - e_{1}^{(-)} \otimes e_{2}^{(-)} \otimes e_{3}^{(0)} \, .
    \label{2-30a}
\eeqar
In (\ref{2-30}), we also define $e_{1}^{(\pm)} \otimes e_{2}^{(h_2)}
\otimes \cdots \otimes e_{m}^{(h_m)} \otimes e_{1}^{(0 )}$ as
\beqar
    e_{1}^{( \pm )} \otimes e_{2}^{(h_2)} \otimes \cdots \otimes e_{m}^{(h_m)} \otimes e_{1}^{(0 )}
    & \equiv & \hf [ e_{1}^{(0)} , e_{1}^{(\pm )} ] \otimes e_{2}^{(h_2 )}
    \otimes \cdots \otimes e_{m}^{(h_m)}
    \nonumber \\
    &=&
    \pm \hf e_{1}^{( \pm )} \otimes e_{2}^{(h_2)} \otimes \cdots \otimes e_{m}^{(h_m)}
    \label{2-30b}
\eeqar
where we implicitly use an antisymmetric property for the indices due to the wedge products.

The trace $\Tr_{R, \ga}$ in the definition (\ref{2-25}) means
a trace over the Chan-Paton factors of the frame fields.
As discussed in \cite{Abe:2009kq},
this trace includes not only a trace over the translational
operators but also that of braid generators.
The latter, a so-called braid trace, is realized by
a sum over permutations of the numbering indices.
Thus the trace $\Tr_{R, \ga}$ over the exponent of (\ref{2-25}) can be
expressed as
\beq
    \Tr_{R, \ga} \Path \sum_{m \ge 2} \oint_{\ga}
    \underbrace{E \wedge \cdots \wedge E}_{m}
    = \sum_{m \ge 2}
    \sum_{\si^\prime \in \S_{m-1}} \oint_{\ga}  E_{1 \si^{\prime}_{2}}
    E_{\si^{\prime}_{2} \si^{\prime}_{3}} \cdots E_{\si^{\prime}_{m} 1}
    \, \om_{1 \si_2} \wedge \om_{\si^{\prime}_{2} \si^{\prime}_{3}} \wedge \cdots \wedge
    \om_{\si^{\prime}_{m} 1}
    \label{2-31}
\eeq
where the sum of $\si^\prime \in \S_{m-1}$ is now taken over the
permutations
$\si^\prime = \left(%
\begin{array}{c}
  2 ~ 3 ~ \cdots ~ m \\
  \si^{\prime}_{2} \si^{\prime}_{3} \cdots \si^{\prime}_{m} \\
\end{array}%
\right)$.

\vskip 5mm
\noindent
\underline{The exponent of $\Theta_{R, \ga}^{(H)} (u, \bu )$}

We now return to the gravitational case.
The above meanings of $\Tr_{R, \ga}$ and $\Path$ can also be applied to (\ref{2-1}).
(Regarding what $\ga$ and an integral around it mean in a gravitational theory,
we shall consider in the next section.)
We can then define a gravitational analog of (\ref{2-31}).
Its explicit form is given by
\beqar
    \nonumber
    && \!\!\!\! \Tr_{R, \ga} \, \Path \oint_{\ga}
    \underbrace{H \wedge H \wedge \cdots \wedge H}_{m}
    \\
    \nonumber
    &=& \!\! \left( 8 \pi G_N \right)^{\frac{m}{2}}
    \Tr_{R , \ga} \oint_{\ga} H_{12} H_{23} \cdots H_{m1}
    ~ \om_{12} \wedge \om_{23} \wedge \cdots \wedge \om_{m1}
    \\
    \nonumber
    &=& \!\!
    \left( 8 \pi G_N \right)^{\frac{m}{2}}
    \left( \frac{1}{2^{m+1}} \right)^{2} \sum_{\si \in \S_{r-1} }
    \sum_{\tau \in \S_{m-r-2} }
    \left(
    \prod_{i=2}^{r} T^{\si_i} \, \prod_{i=r+1}^{m-2} T^{\tau_i}
    \right)
    \\
    \nonumber
    && ~ \times \! \sum_{(h_{11} , h_{2 \si_2} , \cdots , h_{m m} )} \!
    g_{1 1}^{(h_{11})} \otimes g_{2 \si_2}^{(h_{2 \si_2})} \otimes
    g_{3 \si_3}^{(h_{3 \si_3})} \otimes \cdots \otimes g_{r \si_r}^{(h_{r \si_r})}
    \\
    \nonumber
    && \hskip 2.7cm
    \otimes \, g_{r+1 \, \tau_{r+1} }^{(h_{r+1 \, \tau_{r+1}})}
    \otimes g_{r+2 \, \tau_{r+2} }^{(h_{r+2 \, \tau_{r+2}})}
    \otimes \cdots \otimes g_{m-2  \, \tau_{m-2} }^{(h_{m-2 \, \tau_{m-2}})}
    \otimes g_{m-1 \, m-1}^{(h_{m-1 \, m-1})} \otimes g_{m m}^{(h_{m m})}
    \\
    \nonumber
    && ~ \times
    \oint_{\ga} \om_{ 1 2 } \wedge \om_{ 2 3 } \wedge \cdots \wedge \om_{m-1 \, m}
    \wedge \om_{m 1}
    \\
    \nonumber
    && ~ \times
    \oint_{\ga} \om_{ \si_{2} \si_{3}} \wedge \om_{\si_{3} \si_{4}} \wedge \cdots \wedge
    \om_{\si_{r-1} \si_{r}} \wedge \om_{\si_{r} 1}
    \\
    \nonumber
    && \hskip 3.8cm
    \wedge \, \om_{1 \, m-1}  \wedge \om_{m-1 \, \tau_{r+1}} \wedge \om_{\tau_{r+1} \tau_{r+2}}
    \wedge \cdots \wedge \om_{\tau_{m-2} \, m } \wedge \om_{m \si_{2}}
    \\
    &&
    ~ + \, \P (2 3 \cdots m-2)
    \label{2-32}
\eeqar
where the sum of $(h_{11} , h_{2 \si_2} , \cdots , h_{m m} )$ is taken over any
combinations of $h_{i \mu_i} = (++, +-,-+, --)$.
Explicit forms of $T^{\si_i}$'s and $T^{\tau_i}$'s are defined in
(\ref{2-12}) and (\ref{2-13}), respectively.

In (\ref{2-32}), a sum over possible metrics is realized by
two distinct sums over the permutations of
$\si=\left(%
\begin{array}{c}
  2 \cdots r \\
  \si_2 \cdots \si_r \\
\end{array}%
\right)$
and
$\tau=\left(%
\begin{array}{c}
  r+1 \cdots m-2 \\
  \tau_{r+1} \cdots \tau_{m-2} \\
\end{array}%
\right)$. We shall call this set of sums a split sum.
A braid trace, on the other hand, is realized by
$\P(23 \cdots m-2)$ which indicates
the terms obtained by permutations of the overall elements
$\{2,3, \cdots, m-2 \}$.
These realizations reflect the fact that we split the numbering
indices as shown in (\ref{2-3}).
In fact, this feature is pertinent to KLT-inspired graviton amplitudes in general.
In the next section, we treat the numbering indices in a more democratic
manner and consider an alternative definition of
the gravitational holonomy operator (\ref{2-1}).

%%%%%%%%%%%%%%%%%%%%%%%%%%%%%%%%%%%%%%%%%%%%%%%%%%%%%%
%\section{A split sum and a homogeneous sum}
\section{Gravitational holonomy in a squared form}

In this section, we present a main result of this paper.
Namely, we shall obtain an alternative expression for
the exponent of the gravitational holonomy operator, which is different form (\ref{2-32}),
such that we can interpret $\Theta_{R, \ga}^{(H)} (u, \bu )$
as a square of $\Theta_{R, \ga}^{(E)} (u)$.

\vskip 5mm
\noindent
\underline{A split sum and a homogeneous sum}

In the previous section, we have two distinct permutations (\ref{2-8}).
Accordingly, the numbering elements are split into two parts.
Under the ordering conditions,
$\si_2 < \si_3 < \cdots < \si_{r}$ and $\tau_{r+1} < \tau_{r+2}
< \cdots < \tau_{m-2}$, these elements can uniquely be determined.
The braid trace is then realized by
a sum over permutations of the overall elements $\{ 2, 3, \cdots , m-2 \}$.
This sum (or trace) should be taken on top of the split sum, {\it i.e.},
the two distinct sums over $\si$'s and $\tau$'s.
There is another way of calculating the Chan-Paton factor in (\ref{2-32}).
This can be carried out by assigning $\si$'s and $\tau$'s
to the overall elements $\{ 2, 3, \cdots , m-2 \}$ homogeneously.
Namely, the elements of both $\si$'s and $\tau$'s can take any values in
the overall elements, given that they satisfy the ordering conditions.
A primordial form of this alternative expression is first introduced in
the study of graviton amplitudes \cite{Abe:2005se}.
We briefly review its results in the appendix of the present paper.
In the following, we shall use these results in relation to the homogeneous sum
and give an interpretation of $\Theta_{R, \ga}^{(H)}$ as a square of $\Theta_{R, \ga}^{(E)}$.

\noindent
\underline{Symmetry of holonomy operator and characterization of braid trace}

To begin with, we first remind ourselves that the Chan-Paton factor of
(\ref{2-32}) has an $SL (2 ,{\bf C} )$ symmetry.
In the spinor-momenta formalism, this symmetry is relevant to
the Lorentz invariance of the spinor momenta as shown in (\ref{2-19}).
In the Yang-Mills case, the corresponding Chan-Paton factor has a $U(1)$ symmetry.
(Notice that it is a Chan-Paton factor of a Yang-Mills holonomy operator {\it per se},
not that of a gauge field.)
This corresponds to the fact that there is a single type of
permutation, labeled by $\si^\prime$, in the expression (\ref{2-31}).
In terms of the braid trace $\Tr_\ga$, this means that there is a
single loop, say $\ga_1$, that labels the braid trace.
On the other hand, in the gravitational case, we have the $SL (2 ,{\bf C} )$ symmetry.
Thus, as discussed in \cite{Abe:2009kq},
the gravitational braid trace can be characterized by three distinct loops.
These may be chosen by $( \ga_1 , \ga_{m-1},  \ga_m )$
where the indices $(1, m-1, m)$ correspond to the fixed
numbering indices in Figure \ref{fig02}.

There must be a correspondence between the
loops $( \ga_1 , \ga_{m-1} , \ga_m )$ and
the elements of $SL(2, {\bf C})$ algebra, say, a set of
generators $( t^{(+)} , t^{(-)} , t^{(0)} )$.
As discussed in \cite{Abe:2009kq}, a natural way
to realize this correspondence is to assign orderings
to the numbering indices for each of the loops.
We then make the numbering indices in
a descending order for the loop $\ga_{1}$ and in an ascending order
for the loop $\ga_{m-1}$, along with certain orientations of the loops.
The loop $\ga_m$ which corresponds to $t^{(0)}$ does not have
a notion of ordering. Hence, it is natural to think that
the loop $\ga_m$ involves only one numbering element, otherwise we may
have redundant $U(1)$ symmetries.
The gravitational braid trace is therefore essentially
characterized by the ladder generators $t^{(\pm)}$ of $SL(2, {\bf C})$.

\noindent
\underline{Use of the homogeneous sum}

We denote the elements of $\ga_1$ by $\{\si_2 , \si_3 , \cdots ,
\si_{r} \}$ and those of $\ga_{m-1}$ by $\{ \tau_{r+1} , \tau_{r+2} ,
\cdots , \tau_{m-2} \}$ ($2 \le r \le m-3 $).
Then the three disconnected loops can be created by the
three braid diagrams in Figure \ref{fig02}.
In the case of a split sum,
we have split the numbering elements into $\{ 2, 3, \cdots , r \}$
and $\{ r+1 , r+2 , \cdots , m-2 \}$.
Thus, for a specific choice of $r$, this fixes
the choice of the elements for $\si$'s and $\tau$'s.
In the case of a homogeneous sum, however, we assume that
we can choose the numbering elements for $\si$'s (and $\tau$'s)
arbitrarily so that there are $_{m-3}C_{r-1}$ such choices for a fixed $r$.
Using the notations in (\ref{2-3})-(\ref{2-5}) and the results (\ref{A-8})-(\ref{A-10})
in the appendix, we can then write down
a homogeneous version of the expression (\ref{2-32}) as
\beqar
    && \Tr_{R, \ga} \, \Path \oint_{\ga}
    \underbrace{H \wedge H \wedge \cdots \wedge H}_{m}
    \nonumber \\
    &=&
    \left( 8 \pi G_N \right)^{\frac{m}{2}}
    \left( \frac{1}{2^{m+1}} \right)^{2}
    \sum_{\{\si,\tau \}=\{2,3, \cdots, m-2 \}}
    \, \Biggl[ ~
    \C(\mu_1 \mu_2 \cdots \mu_m )
    \nonumber \\
    && \!\!\! \times
    \left( \,
    \prod_{i=1}^{m} T^{\la_i} \!\!
    \sum_{( h_{\mu_1 \la_1} , h_{\mu_2 \la_2}, \cdots, h_{\mu_m \la_m} ) } \!\!
    g_{\mu_1 \la_1}^{(h_{\mu_1 \la_1})} \otimes
    g_{\mu_2 \la_2}^{(h_{\mu_2 \la_2})} \otimes
    \cdots \otimes g_{\mu_m \la_m}^{(h_{\mu_m \la_m})}
    ~ \C(\la_1 \la_2 \cdots \la_m ) +  \P(\si |  \tau ) \right)
    \nonumber \\
    &&
     + ~ \P(\si  |  \tau )
    ~ \Biggr]
    \label{3-1}
\eeqar
where
\beqar
    \C(\mu_1 \mu_2 \cdots \mu_m )
    &=& \C(1 \si_2 \cdots \si_r \tau_{r+1} \cdots \tau_{m-2} \, m-1 \, m )
    \nonumber \\
    &=&
    \oint_\ga \om_{1 \si_2} \wedge \om_{ \si_2 \si_3 } \wedge \cdots \wedge
    \om_{m-1 \, m} \wedge \om_{m 1}
    \label{3-2}
\eeqar
and the same for $\C(\la_1 \la_2 \cdots \la_m ) =
\C(\si_2 \cdots \si_r \, 1 \, m-1 \, \tau_{r+1} \cdots \tau_{m-2} \, m)$.
The sum of $\{\si,\tau \}$ in (\ref{3-1}) is taken over the all possible combinations for
the elements $\{ \si_2 , \cdots , \si_r , \tau_{r+1} , \cdots , \tau_{m-2} \}$
such that the ordering conditions
$\si_2 < \cdots < \si_r$ and $\tau_{r+1} < \cdots < \tau_{m-2} $ are preserved.
As in the case of (\ref{2-32}), the
sum of $(h_{\mu_1 \la_1} , h_{\mu_2 \la_2}, \cdots, h_{\mu_m \la_m} )$ is taken over any
combinations of $h_{\mu_i \la_i} = (++, +-,-+, --)$ with $i=1,2,\cdots, m$.
Notice that the factor $\prod_{i=1}^{m} T^{\la_i}$
is determined only by the permutations $\si$ and $\tau$.
Thus it is also equal to $\prod_{i=1}^{m} T^{\mu_i}$.
In other words, this product sum is
in one-to-one correspondence with the braid diagrams in Figure \ref{fig02},
and is therefore uniquely determined once we choose the permutations $\si$ and $\tau$.
The symbol $\P ( \si | \tau )$ in (\ref{3-1}) denotes the terms obtained by
the permutations of $\si$'s and $\tau$'s, {\it i.e.},
\beq
    \P (\si | \tau ) = \P ( \si_2 \si_3 \cdots \si_r ) \times \P ( \tau_{r+1}
    \tau_{r+2} \cdots \tau_{m-2} )
    \label{3-3}
\eeq
where, as in (\ref{2-32}), $\P ( \si_2 \si_3 \cdots \si_r )$ denotes
terms obtained by permutations of $\si$'s, and the same for $\P (\tau_{r+1}
\tau_{r+2} \cdots \tau_{m-2} )$.

\noindent
\underline{A double braid-trace and the squared form}

As in the previous case, $\P ( \si | \tau )$ can be
regarded as a realization of a braid trace. The double appearance of
$\P ( \si | \tau )$ then suggests that the braid trace over gravitons can be
replaced by a double braid-trace over frame fields.
This interpretation is in accord with the idea that the graviton
is describable in terms of a product of frame fields
even at the level of comprehensive field operators.
As we shall see in the next section,
the double appearance of $\P (\si | \tau )$ also supports
the use of functional derivatives with respect to
the frame-field operators (or source functions, to be precise)
in obtaining graviton amplitudes generated by
the gravitational holonomy operator.
Notice that $\P (\si | \tau )$'s appear before taking the
homogeneous sum of $\{ \si , \tau \}$.
{\it Thus, the eventual expression of (\ref{3-1}) is independent of
the choices of $\si$'s and $\tau$'s, but the squared structure appears
inside the homogeneous sum and, in this respect, we need to label
the indices of frame fields by $( \si , \tau)$ or equivalently
by $(\mu , \la )$.
}

Motivated by these considerations, we now introduce a new notation:
\beqar
    && \Tr_{R_\mu , \ga_{\si | \tau }} \, \Path \oint_{\ga_{\si | \tau } }
    \underbrace{E \wedge E \wedge \cdots \wedge E}_{m}
    \nonumber \\
    &=&
    \Tr_{R_\mu , \ga_{\si | \tau } } \, \oint_{\ga_{\si | \tau } }
    E_{ \mu_1 \mu_2} E_{\mu_2 \mu_3} \cdots  E_{\mu_m \mu_1}
    \, \om_{\mu_1 \mu_2} \wedge \om_{ \mu_2 \mu_3 } \wedge \cdots \wedge \om_{\mu_m \mu_1}
    \nonumber \\
    &=&
    E_{\mu_1 \mu_2} E_{\mu_2 \mu_3} \cdots  E_{\mu_m \mu_1} ~
    \C( \mu_1 \mu_2 \cdots \mu_m )
    + \P ( \si | \tau )
    \label{3-4}
\eeqar
where we denote the closed path by $\ga_{\si | \tau }$ to indicate
that the permutations over the numbering indices are separately taken for
$\si$'s and $\tau$'s.
We also label the representation of the algebra of a braid group by $R_\mu $,
which reflects that the comprehensive frame fields are labeled by $\mu_i$
($i = 1,2,\cdots , m$) in the above expression.

Using the notation (\ref{3-4}), we can rewrite (\ref{3-1}) as
\beqar
    && \Tr_{R, \ga} \, \Path \oint_{\ga}
    \underbrace{H \wedge H \wedge \cdots \wedge H}_{m}
    \nonumber \\
    &=&
    \left( 8 \pi G_N \right)^{\frac{m}{2}} \!
    \sum_{\{\si,\tau \}=\{2,3, \cdots, m-2 \}} \!
    \nonumber \\
    &&
    \left\langle \Tr_{R_\mu , \ga} \, \Path \oint_{\ga }
    \underbrace{E \wedge E \wedge \cdots \wedge E}_{m}
    \, \cdot \,
    \Tr_{R_\la , \ga} \, \Path \oint_{\ga }
    \underbrace{E \wedge E \wedge \cdots \wedge E}_{m}
    \right \rangle_{\ga = \ga_{\si | \tau }}
    \nonumber \\
    &=&
    \left( 8 \pi G_N \right)^{\frac{m}{2}}
    \left( \frac{1}{2^{m+1}} \right)^{2}
    \sum_{\{\si,\tau \}=\{2,3, \cdots, m-2 \}}
    \, \Tr_{R_\mu , \ga } \, \Biggl[ \,
    \C(\mu_1 \mu_2 \cdots \mu_m )
    \nonumber \\
    &&
    \times \,
    \Tr_{R_\la , \ga} \, \biggl[ ~
    \prod_{i=1}^{m} T^{\la_i} \!\!
    \sum_{( h_{\mu_1 \la_1} , \cdots, h_{\mu_m \la_m} ) } \!
    g_{\mu_1 \la_1}^{(h_{\mu_1 \la_1})} \otimes
    \cdots \otimes g_{\mu_m \la_m}^{(h_{\mu_m \la_m})}
    ~ \C(\la_1 \la_2 \cdots \la_m ) ~
    \biggr] \,
    \Biggr]_{\ga = \ga_{\si | \tau }}
    \label{3-5}
\eeqar
where the bracket in the middle denotes a set of products
between Chan-Paton factors of the frame fields, as defined in (\ref{2-24}).
The set of products take a form of $\prod_{i=1}^{m} T^{\la_i}$ in the last line.
As mentioned below (\ref{3-2}), this factor is dependent only on the choice
of $(\si , \tau )$, {\it i.e.},
\beq
    \prod_{i=1}^{m} T^{\la_i} \, = \,
    \prod_{i=2}^{r} T^{\si_i} \prod_{i=r+1}^{m-2} T^{\tau_i}
    \, = \, \prod_{i=1}^{m} T^{\mu_i} \, .
    \label{3-8a}
\eeq
This reflects the fact that the holonomy operator inherently
describes a multi-particle system so that its Chan-Paton factor
depends on comprehensive information about all the particles.
The information is given by a specific permutation of particles
in a form of an irreducible representation of the braid diagrams.
As analyzed in \cite{Abe:2009kq}, the factor (\ref{3-8a}) is indeed
in one-to-one correspondence with the braid diagrams in Figure \ref{fig02},
once we impose irreducibility up to certain Reidemeister moves of the strands.
The specific choice of $\la_i$, in relation to $\mu_i$,
has been made such that we have non-vanishing Chan-Paton factors.
In this sense, the indices $\mu_i$ and $\la_i$
labels the (irreducible) representation of the diagrams.
This means that, before carrying out the homogeneous sum,
the gravitational holonomy operator can and should be labeled by
the representation $R_{\mu \la}$.
Notice that in the Yang-Mills case, we do not have to consider
the product of Chan-Paton factors such as (\ref{3-8a}).
Thus, in taking the homogeneous sum, a representation of the algebra of
a braid group, or a representation of the Iwahori-Hecke algebra,
for the frame-field holonomy operator can be labeled by a single index.
This explains the notations $R_\mu$ and $R_{\la}$ in (\ref{3-5}).

{\it Using the expression (\ref{3-5}), we can then express
the gravitational holonomy operator as
a homogeneous sum over squares of the frame-field holonomy operator:}
\beqar
    \Theta_{R, \ga}^{(H)} ( u, \bu ) & = &
    \sum_{\{\si,\tau \}}
    \Theta_{R_{\mu \la}, \ga_{\si | \tau }}^{(H)} ( u, \bu ) ~ = ~
    \sum_{\{\si,\tau \}}
    \left\langle \Theta_{R_\mu , \ga  }^{(E)} ( u ) \cdot
    \Theta_{R_\la , \ga  }^{(E)} ( u )
    \right \rangle_{\ga = \ga_{\si | \tau }}
    \label{3-6} \\
    \Theta_{R_\mu , \ga_{\si | \tau }  }^{(E)} ( u )
    &=&
    \Tr_{R_\mu , \ga_{\si | \tau }  } \Path \exp \left[
    \sum_{m \ge 2} \oint_{\ga_{\si | \tau }}
    \underbrace{E \wedge E \wedge \cdots \wedge E}_{m}
    \right]
    \label{3-7} \\
    \Theta_{R_\la , \ga_{\si | \tau }  }^{(E)} ( u )
    &=&
    \Tr_{R_\la , \ga_{\si | \tau }  } \Path \exp \left[
    \sum_{m \ge 2} \oint_{\ga_{\si | \tau }}
    \underbrace{E \wedge E \wedge \cdots \wedge E}_{m}
    \right]
    \label{3-8}
\eeqar
where we make the coupling constant absorbed into
each of the frame-field operators.
We specify the representation of the frame-field holonomy operators
by $R_\mu$ and $R_\la$.
This corresponds to the fact that the exponent of
$\Theta_{R, \ga}^{(H)} ( u, \bu )$ is given by the expression (\ref{3-5}).
The homogeneous sum of $\{\si,\tau \}$ is taken over the all possible combinations for
the elements $\{ \si_2 , \cdots , \si_r , \tau_{r+1} , \cdots , \tau_{m-2} \}
= \{ 2,3, \cdots , m-2 \}$ such that the ordering conditions $\si_2 < \si_3 < \cdots < \si_r$
and $\tau_{r+1} < \tau_{r+2} < \cdots < \tau_{m-2} $ are preserved.

\noindent
\underline{The homogeneous sum: a sum over $(k,l)$-shuffles}

We now briefly discuss how the homogeneous sum appears naturally
in the framework of holonomy formalism.
The factor of $\C ( \mu_1 \mu_2 \cdots \mu_m)$ in (\ref{3-2}) gives
an iterated (loop) integral over a series of the logarithmic one-forms.
Generally, a product of iterated integrals can be defined as \cite{Kohono:2009bk}
\beq
    \int_{{\tilde{\ga}}_k} \om_{1} \om_{2} \cdots \om_{k}
    \,
    \int_{{\tilde{\ga}}_l} \om_{k+1} \om_{k+2} \cdots \om_{k+l}
    \, = \,
    \sum_{\tilde{\si} \in S_{k, l} }
    \int_{{\tilde{\ga}}_{k+l}} \om_{\tilde{\si}_1}
    \om_{\tilde{\si}_2} \cdots \om_{\tilde{\si}_{k+l}}
    \label{3-9}
\eeq
where $\om_1 , \om_2 , \cdots , \om_{k+l}$ are arbitrary differential one-forms.
The symbol $\tilde{\ga}_n$ denotes an open path in ${\bf C}^n$.
The sum of the permutations $\tilde{\si} \in S_{k, l}$
is taken over the so-called $(k , l)$-shuffles $S_{k, l}$ that satisfy
the ordering conditions:
\beq
    \begin{array}{c}
    \tilde{\si}_{1} < \tilde{\si}_{2} < \cdots < \tilde{\si}_{k} \, , \\
    \tilde{\si}_{k+1 } < \tilde{\si}_{k+2} < \cdots < \tilde{\si}_{k+l} \, . \\
    \end{array}
    \label{3-10}
\eeq
The sum of $\tilde{\si} \in S_{k, l}$ is therefore essentially the same as the homogeneous sum.
Applying the relation (\ref{3-9}) to loop integrals along $( \ga_1 , \ga_{m-1} , \ga_{m} )$
that we have defined in the beginning of this section, we can then obtain an expression
\beqar
    &&
    \sum_{\{\si,\tau \}} \oint_{\ga_{\si | \tau } }
    \om_{\mu_1} \om_{ \mu_2 } \cdots \om_{\mu_m }
    \nonumber \\
    &=&
    \oint_{\ga_1} \om_1 \om_2 \cdots \om_r \,
    \oint_{\ga_{m-1}} \om_{r+1} \om_{r+2} \cdots \om_{m-1} \,
    \oint_{\ga_m} \om_m
    \nonumber \\
    &=&
    \oint_{\ga_1}  \om_2 \om_3 \cdots \om_r \om_1 \,
    \oint_{\ga_{m-1}} \om_{m-1} \om_{r+1} \om_{r+2} \cdots \om_{m-2} \,
    \oint_{\ga_m} \om_m
    \nonumber \\
    &=&
    \sum_{\{\si,\tau \}} \oint_{\ga_{\si | \tau } }
    \om_{\la_1} \om_{ \la_2 } \cdots \om_{\la_m }
    \label{3-11}
\eeqar
where we use the cyclic property of the loop integrals
along $\ga_1$ and $\ga_{m-1}$.
The sum of $\{\si,\tau \}$ denotes the homogeneous sum, being the
same as the one defined in (\ref{3-1}).
As mentioned earlier, the overall path
$\ga  = \ga_{\si | \tau}$ is decomposed into three closed paths
$ ( \ga_1 , \ga_{m-1} , \ga_m )$.
By identifying $\om_i$ as the logarithmic one-form $\om_{i \, i+1}$
in (\ref{3-11}), we can then obtain the relation
\beq
    \sum_{\{\si,\tau \}}
    \C ( \mu_1 \mu_2 \cdots \mu_m )
    \, = \,
    \sum_{\{\si,\tau \}}
    \C ( \la_1 \la_2 \cdots \la_m ) \, .
    \label{3-12}
\eeq
This equation means that the factors of
$\C ( \mu_1 \mu_2 \cdots \mu_m )$ and
$\C ( \la_1 \la_2 \cdots \la_m )$ are equivalent under the
homogeneous sum or the sum over the $(r-1 , m-2-r)$-shuffles.
Thus, in this sense, the subset of the gravitational holonomy operator,
denoted by $\Theta_{R_{\mu \la}, \ga_{\si | \tau }}^{(H)} ( u, \bu ) $ in (\ref{3-6}),
can be interpreted as a square of the same theory.

\noindent
\underline{General covariance and diffeomorphism}

Lastly, as a summary of this section, we now consider
some physical aspects of the squared expression (\ref{3-1}) or (\ref{3-5}).
There are essentially two sums to be taken in the holonomy formalism of gravity.
As emphasized in \cite{Abe:2009kq}, these are given by the following two sums:
\begin{enumerate}
  \item a sum over all possible metrics that guarantees general covariance of the theory; and
  \item a sum over permutations of the numbering elements, or a braid trace, that is
  necessary for diffeomorphism invariance.
\end{enumerate}
For the original expression (\ref{2-32}), as discussed in the previous section,
the former sum is realized by the split sum
$\displaystyle \sum_{\si \in \S_{r-1}} \sum_{\tau \in \S_{m-r-1}}$ and
the latter is represented by the terms of $\P ( 2 3 \cdots m-2 )$.
In this section, we have shown that the double appearance
of $\P (\si | \tau )$ in (\ref{3-1}) can be interpreted as a double braid-trace in (\ref{3-5}).
Thus, in the squared expression, the braid trace is realized by
the double-permutation terms, which we denote here as $\P (\si | \tau ) + \P (\si | \tau )$,
while the sum over metrics is realized by the homogeneous sum
$\displaystyle \sum_{ \{\si,\tau \}=\{2,3, \cdots, m-2 \} }$.
Therefore, for either case, we can make
physically clear interpretations to the summations that appear
in the expressions of the gravitational holonomy operator.
These interpretations are summarized in Table \ref{table01}.

\begin{table}[htbp]
\begin{center}
\begin{tabular}{|r||c|c|}
  \hline
  Sum over metrics & split sum & homogeneous sum \\
  (general covariance)
  & \small{$\displaystyle \sum_{\si \in \S_{r-1} } \sum_{ \tau \in \S_{m-r-1} }$ }
  & \small{$\displaystyle \sum_{ \{\si,\tau \}=\{2,3, \cdots, m-2 \} }$ }
  \\
  \hline
  Braid trace  & single-permutation terms & double-permutation terms \\
  (diffeomorphism)
  & $\P ( 23 \cdots m-2 )$ & $\P( \si | \tau ) + \P (\si | \tau )$ \\
  \hline
  Quantities of interest & gravitons & frame fields  \\
  \hline
  Relevant expression & (\ref{2-32}) & (\ref{3-1}), (\ref{3-5})  \\
  \hline
  Gravitational theory & as a gauge theory & as a square of gauge theory  \\
  \hline
\end{tabular}
\caption{Interpretation of a sum over metrics and a braid trace in
expressions of the gravitational holonomy operator}
\label{table01}
\end{center}
\end{table}

\section{S-matrix functionals for graviton amplitudes}

So far, we have discussed how the gravitational holonomy operator can be
expressed as a square of the frame-field holonomy operator.
In this section, we utilize the new expression to obtain
an S-matrix functional for graviton amplitudes.
For this purpose, we first review how we obtain
an S-matrix functional for the maximally helicity violating (MHV)
graviton amplitudes in the split-sum case.
We then consider the homogeneous-sum case and show that
the MHV S-matrix functional
can also be described in terms of a supersymmetric version of the operator (\ref{3-6}).
For the completion of the analysis, we shall also consider S-matrix functionals
for non-MHV amplitudes in general.

\noindent
\underline{The split-sum case}
%\subsection{The split-sum case}

In the holonomy formalism, physical information is embedded in
the operator $g_{i}^{(h_{i \mu_i})}$ in (\ref{2-5}).
This operator is in a momentum-space representation.
Let $x$ be the four-dimensional spacetime coordinate.
In an $x$-space representation, the operator is then expressed as
\beq
    g_{i}^{(h_{i \mu_i} )} (x) \, = \, \int d \mu ( p_i ) ~ g_{i}^{(h_{i \mu_i} )}
    ~ e^{i x \cdot p_i }
    \label{4-1}
\eeq
where $d \mu  (p_i)$ denotes a four-dimensional Lorentz invariant measure, known as
the Nair measure.

It is known that the most convenient prescription to
an S-matrix functional for the MHV amplitudes is to supersymmetrize
the operator (\ref{4-1}).
In the present case, we consider an ${\cal N}=8$ extended supersymmetry.
The relevant Grassmann variables are expressed as
$\th^{\al}_{A}$, with $A = 1, 2$ and $\al = 1,2, \cdots , 8$.
In the spinor-momenta formalism, it is convenient to introduce
the ``projected'' Grassmann variables:
\beq
    \xi^\al \, = \, \th_{A}^{\al} u^A \, ~~~ (\al = 1,2, \cdots , 8).
    \label{4-2}
\eeq
For the later convenience, we further split the index $\al$ into two parts:
\beq
    \begin{array}{c}
     \al \, = \, ( \al_1 , \al_2 ), \\
     \al_1 = 1,2,3,4, ~~~ \al_2 = 5,6,7,8 . \\
    \end{array}
    \label{4-3}
\eeq
We can then write down a supersymmetrization of (\ref{4-1}) as
\beqar
    g_{i}^{(\hat{h}_{i \mu_i})} (x, \th)
    & = &
    \left. \int d\mu (p_i) ~ g_{i}^{(\hat{h}_{i \mu_i})} ( \xi_{i} )
    ~  e^{ i x \cdot p_{i} }
    \right|_{\xi_{i}^{\al} = \th_{A}^{\al} u_{i}^{A} }
    \label{4-4}
    \\
    g_{i}^{(\hat{h}_{i \mu_i})} ( \xi_{i} ) &=&
    T^{\mu_i} \, g_{i \mu_i}^{(\hat{h}_{i \mu_i})} ( \xi_{i} )
    \, = \,
    T^{\mu_i} \, e_{i}^{(\hat{h}_{i}) a} ( \xi_i ) \,
    e_{\mu_i}^{(\hat{h}_{\mu_i}) a} ( \xi_i )
    \label{4-5}
\eeqar
where $g_{i}^{(\hat{h}_{i \mu_i})} ( \xi_{i} )$
in the second equation can be considered as a supersymmetrization of
the graviton operator $g_{i}^{( h_{i \mu_i } )}$ defined in (\ref{2-5}).
Accordingly, $e_{i}^{(\hat{h}_{i}) a} (\xi_i)$ ($a=0,1,2,3$) correspond
to a supersymmetric version of the frame fields $e_{i}^{( h_i )}$ in (\ref{2-16})
and are defined as
\beqar
    \nonumber
    e_{i}^{(+)a} (\xi_i) &=& e_{i}^{(+) a} \\ \nonumber
    e_{i}^{\left( + \hf \right) a} (\xi_i) &=& \xi_{i}^{\al_1}
    \, e_{i \, \al_1}^{ \left( + \hf \right) a} \\
    e_{i}^{(0)a} (\xi_i) &=& \hf \xi_{i}^{\al_1} \xi_{i}^{\bt_1} \, e_{i \, \al_1 \bt_1}^{(0)a}
    \label{4-6}
    \\ \nonumber
    e_{i}^{\left(- \hf \right)a} (\xi_i) &=& \frac{1}{3!} \xi_{i}^{\al_1}\xi_{i}^{\bt_1}
    \xi_{i}^{\ga_1}
    \ep_{\al_1 \bt_1 \ga_1 \del_1} \, {e_{i}^{ \del_1}}^{ \left( - \hf \right) a}
    \\ \nonumber
    e_{i}^{(-)a} (\xi_i) &=& \xi_{i}^{1} \xi_{i}^{2} \xi_{i}^{3} \xi_{i}^{4} \, e_{i}^{(-)a}
\eeqar
where each of $\al_1 ,  \bt_1 , \cdots$ takes a value of 1, 2, 3 or 4.
Similarly, $e_{\mu_i}^{(\hat{h}_{\mu_i}) a} (\xi_i)$'s are defined as
\beqar
    \nonumber
    e_{\mu_i}^{(+)a} (\xi_i) &=& e_{\mu_i}^{(+) a} \\ \nonumber
    e_{\mu_i}^{\left( + \hf \right) a} (\xi_i) &=& \xi_{i}^{\al_2}
    \, e_{\mu_i \, \al_2}^{ \left( + \hf \right) a} \\
    e_{\mu_i}^{(0)a} (\xi_i) &=& \hf \xi_{i}^{\al_2} \xi_{i}^{\bt_2} \,
    e_{\mu_i \, \al_2 \bt_2}^{(0)a}
    \label{4-7}
    \\ \nonumber
    e_{\mu_i}^{\left(- \hf \right)a} (\xi_i) &=& \frac{1}{3!} \xi_{i}^{\al_2}\xi_{i}^{\bt_2}
    \xi_{i}^{\ga_2}
    \ep_{\al_2 \bt_2 \ga_2 \del_2} \, {e_{\mu_i}^{ \del_2}}^{ \left( - \hf \right) a}
    \\ \nonumber
    e_{\mu_i}^{(-)a} (\xi_i) &=& \xi_{i}^{5} \xi_{i}^{6} \xi_{i}^{7} \xi_{i}^{8} \,
    e_{\mu_i}^{(-)a}
\eeqar
where each of $\al_2 ,  \bt_2 , \cdots$ takes a value of 5, 6, 7 or 8.
Notice that either $\hat{h}_i$ or $\hat{h}_{\mu_i}$ represents a
helicity of an $\N = 4$ supersymmetric frame field.
The symbol $\hat{h}_{i \mu_i}$ then denotes a supersymmetrization of
$h_{i \mu_i} \equiv h_i h_{\mu_i}$, {\it i.e.},
$\hat{h}_{i \mu_i} \equiv \hat{h}_{i} \hat{h}_{\mu_i}$.
We use $\xi_{i}^{\al_2}$'s, rather than $\xi_{\mu_i}^{\al_2}$'s, in (\ref{4-7}).
This comes from the fact that we interpret the graviton (\ref{4-4}) as a point-like
operator in $\N = 8$ chiral superspace.
Alternatively, we can interpret $\xi_i$ as chiral superpartners of the tangent-space
coordinate $x_a$ ($a=0,1,2,3$), with spacetime not being supersymmetrized.

{\it
A supersymmetric gravitational holonomy operator
$\Theta_{R , \ga}^{(H)} ( u , \bu ; x , \th)$
is then  defined by substitution of
$g_{i}^{(\hat{h}_{i \mu_i})} (x, \th)$
into $g_{i}^{(h_{i \mu_i} )}$ in (\ref{2-2}).
}

Using the supersymmetric holonomy operator
$\Theta_{R , \ga}^{(H)} ( u , \bu ; x , \th)$,
we can define an S-matrix functional for the MHV graviton amplitudes as
\beq
    \F_{MHV} \left[ g_{i \mu_i}^{(h_{i \mu_i})} \right]
    \, = \, \exp \left[ \frac{i}{8 \pi G_N} \int d^4 x \, d^{16} \th
    ~ \Theta_{R, \ga}^{(H)} (u, \bu ; x ,\th) \right]
    \label{4-8}
\eeq
where $g_{i \mu_i}^{(h_{i \mu_i})}$ $(i = 1,2, \cdots )$ denotes an operator
or a source function associated with the expression
$g_{i}^{(h_{i \mu_i})} = T^{\mu_i}  g_{i \mu_i}^{( h_{i \mu_i} )}$.

Now, from the general formula for graviton amplitudes (\ref{A-8}) and (\ref{A-9})
in the appendix, we find that the MHV graviton amplitudes can be expressed as follows.
\beqar
    \M^{(s_{--}  t_{--} )}_{MHV} (u, \bu)
    & = &  i ( 8 \pi G_N )^{\frac{n}{2} - 1} (-1)^{n+1}
    \, (2 \pi)^4 \del^{(4)} \left( \sum_{i=1}^{n} p_i \right)
    \widehat{M}^{(s_{--}  t_{--} )}_{MHV}
    (u, \bu)
    \label{4-9}\\
    \widehat{M}^{(s_{--}  t_{--} )}_{MHV} (u, \bu)
    & = &  \sum_{\si \in \S_{r-1}} \sum_{\tau \in \S_{n-r-2}}
    \, \prod_{i=2}^{r} T^{\si_i} \, \prod_{i=r+1}^{n-1} T^{\tau_i}
    ~ \widehat{C}^{(s_{-}  t_{-} )}_{MHV} (1 2 \cdots n)
    \nonumber \\
    && \, \times \, \widehat{C}^{(s_{-}  t_{-} )}_{MHV} (\si_2 \si_3 \cdots \si_{r} \, 1 \, n-1 \,
    \tau_{r+1} \tau_{r+2} \cdots \tau_{n-2} \, n)
    \nonumber \\
    && \, + \, \P (23 \cdots n-2)
    \label{4-10}\\
    \widehat{C}_{MHV}^{ (s_{-} t_{-})} (1 2 \cdots n)
    & = &
    \frac{ (u_s u_t )^4}{ ( u_1 u_{2})( u_{2} u_{3})
    \cdots (u_{n} u_1 )}
    \label{4-11}
\eeqar
where we label the two negative-helicity gravitons by $(s_{--}  t_{--} )$,
with the rest of gravitons having helicity $++ (= +2)$.
From (\ref{4-8}) and (\ref{4-10}), we find that the
MHV graviton amplitudes $\widehat{M}_{MHV}^{(s_{--} t_{--} )} (u, \bu)$
are indeed generated by $\F_{MHV} \left[ g_{i \mu_i}^{(h_{i \mu_i})} \right]$ as
\beqar
    \nonumber
    &&
    \frac{\del}{\del g_{1 \mu_1}^{(++)}( x )} \otimes
    \cdots \otimes \frac{\del}{\del g_{s \mu_s}^{(--) }( x )} \otimes
    \cdots
    \\
    \nonumber
    &&
    ~~~~~~~~~~~~
    \left.
    \cdots \otimes \frac{\del}{\del g_{t \mu_t}^{(--) }( x )} \otimes
    \cdots \otimes \frac{\del}{\del g_{n \mu_n}^{(++)}( x )}
    ~ \F_{MHV} \left[ g_{i \mu_i}^{(h_{i \mu_i})} \right]
     \right|_{g_{i \mu_i}^{(h_{i \mu_i})} ( x ) =0}
    \\
    & = &
    i ( 8 \pi G_N )^{\frac{n}{2}-1} \, \widehat{M}_{MHV}^{(s_{--} t_{--})} (u , \bu)
    \label{4-12}
\eeqar
where we use the result (\ref{2-32}) and the Grassman integral
\beq
    \left.
    \int d^{16} \th \, \prod_{\al = 1}^{8} \xi_{s}^{\al}
    \, \prod_{\bt = 1}^{8} \xi_{t}^{\bt} \right|_{\xi_{i}^{\al} = \th_{A}^{\al} u_{i}^{A}}
     =  \, ( u_s u_t )^8  \, .
    \label{4-13}
\eeq
In obtaining (\ref{4-12}), we also use the normalization relation
\beq
    \oint_\ga d(u_{1} u_{2}) \wedge d(u_{2} u_{3})
    \wedge \cdots \wedge d (u_{n} u_{1}) = 2^{n+1}
    \label{4-14}
\eeq
for the spinor momenta.
Under a permutation of the numbering indices, a sign factor arises in the
above expression. We disregard this sign factor since physical quantities
are given by the square of the amplitudes.
As discussed below (\ref{3-4}),
we can incorporate the information on permutations into the closed path $\ga$
on $\C =  {\bf C}^{n} / \S_{n}$. Thus, we may make this
sign factor absorbed into the above normalization.
Notice that only the MHV-type helicity configurations are survived in
the above calculation (\ref{4-12}).
The rest of the helicity configurations are prohibited due to
the Grassmann integral (\ref{4-13}).

The MHV amplitude (\ref{4-9}) is expressed in a momentum-space representation.
In an $x$-space representation, this can be written as
\beq
    \M^{(s_{--}  t_{--} )}_{MHV} (x) \, = \,
    \prod_{i=1}^{n} \int d\mu (p_i ) \,
    \M^{(s_{--}  t_{--} )}_{MHV} (u, \bu) \, .
    \label{4-15}
\eeq
In terms of the S-matrix functional (\ref{4-8}), this MHV amplitude can also be generated as
\beqar
    \nonumber
    &&
    \left.
    \frac{\del}{\del g_{1 \mu_1}^{(++)}} \otimes
    \cdots \otimes \frac{\del}{\del g_{s \mu_s}^{(--) }} \otimes
    \cdots
    \cdots \otimes \frac{\del}{\del g_{t \mu_t}^{(--) }} \otimes
    \cdots \otimes \frac{\del}{\del g_{n \mu_n}^{(++)}}
    ~ \F_{MHV} \left[ g_{i \mu_i}^{(h_{i \mu_i})} \right]
     \right|_{g_{i \mu_i}^{(h_{i \mu_i})} =0}
    \\
    & = &
    (-1)^{n+1} \, \M_{MHV}^{(s_{--} t_{--})} (x)
    \label{4-16}
\eeqar
where, again, the sign factor $(-1)^{n+1}$ may be irrelevant to physical observables.
Notice that in the above calculation
the momentum-conservation delta function in (\ref{4-9}) naturally arises.

\noindent
\underline{The homogeneous-sum case}
%\subsection{The homogeneous-sum case}

We now consider an alternative expression for the MHV S-matrix functional
by use of the expression (\ref{3-6}) where the homogeneous sum appears.
As summarized in Table \ref{table01}, the physical quantities of interest
in this case are the frame fields rather than the gravitons.
Consequently, the gravitational theory is now given by
a ``square'' of $\N = 4$ theory for the frame fields.
We then have two types of Grassmann variables:
\beqar
    \xi_{\mu_i}^{\al} &=& \th_{A}^{\al} u_{\mu_i}^{A} ~~~~ ( \al = 1,2,3,4; \, A = 1,2 ),
    \label{4-17}\\
    \eta_{\la_i}^{\bt} &=& {\th^\prime}_{B}^{\bt} u_{\la_i}^{B} ~~~~
    ( \bt = 1,2,3,4; \, B = 1,2 ).
    \label{4-18}
\eeqar
One may find that the use of indices $\mu_i$ and $\la_i$ is redundant.
As emphasized in the previous section, however, the squared structure
appears before taking the homogeneous sum of $\{ \si , \tau \}$.
Thus labeling the numbering indices by $\mu_i$ and $\la_i$ is appropriate for our purpose.
Of course, eventually the homogeneous sum is taken so that the final
form is independent of the choice of $(\mu , \la)$ or that of $( \si , \tau )$.

Using the above Grassmann variables, we can define supersymmetric operators for
the frame fields:
\beqar
    e_{\mu_i}^{( \hat{h}_{\mu_i} ) } ( x , \th ) &=&
    \left. \int d \mu (p_{\mu_i} )
    ~ e_{ \mu_{i} }^{( \hat{h}_{\mu_i} ) } (  \xi_{\mu_i} )
    ~  e^{ i x \cdot p_{\mu_i} }
    \right|_{\xi_{\mu_i}^{\al} = \th_{A}^{\al} u_{\mu_i}^{A} }
    \label{4-19} \\
    e_{\la_i}^{( \hat{h}_{\la_i} ) } ( x^\prime , \th^\prime ) &=&
    \left. \int d \mu ( p_{\la_i} )
    ~ e_{\la_i }^{( \hat{h}_{\la_i} ) } ( \eta_{\la_i} )
    ~  e^{ i x^\prime \cdot p_{\la_i} }
    \right|_{\eta_{\la_i}^{\bt} = {\th^\prime}_{B}^{\bt} u_{\la_i}^{B} }
    \label{4-20}
\eeqar
where $d \mu ( p_{\mu_i })$ and $d \mu ( p_{\la_i })$
denote the Nair measures for
$p_{\mu_i}^{A \Ad}= u^{A}_{\mu_i} \bu_{\mu_i}^{\Ad}$ and
$p_{\la_i}^{B \Bd}= u^{B}_{\la_i} \bu_{\la_i}^{\Bd}$, respectively.
As shown in (\ref{2-22}), the spacetime coordinates $x_{A \Ad}$, $x^{\prime}_{B \Bd}$
are defined in terms of twistor-space variables:
\beqar
    v_{\mu_{i} \Ad} &=& x_{A \Ad} u_{\mu_i}^{A} \, ,
    \label{4-21}\\
    v^{\prime}_{\la_{i} \Bd} &=& x^{\prime}_{B \Bd} u_{\la_i}^{B}
    \label{4-22}
\eeqar
where $x_{A \Ad}$ and $x^{\prime}_{B \Bd}$
are two distinct coordinates but $u_{\mu_i}^{A}$ and $u_{\la_i}^{B}$
are those spinor momenta that are defined on the same physical configuration space
$\C = {\bf C}^{n}/ \S_n$.
In the holonomy formalism, physical variables are given by
the spinor momenta. Thus the emergence of two distinct
spacetimes $x$, $x^{\prime}$
is possible but it does seem unnatural in modeling a physical theory.
In the following, we shall consider a gravitational theory such that
$x^\prime$-dependence becomes immaterial.
{\it Our strategy is to define a supersymmetric
gravitational holonomy operator
$\Theta_{R , \ga}^{(H)} ( u , \bu ; x , \th , x^\prime , \th^\prime )$, which
is analogous to the above $\Theta_{R , \ga}^{(H)} ( u , \bu ; x , \th  )$,
and obtain an MHV S-matrix functional from it by integrating out
the $x^\prime$-dependence.}
As we shall see later, it turns out that this construction is
also suitable for the generation of the non-MHV amplitudes.

The frame-field operators (\ref{4-19}) and (\ref{4-20})
are analogs of the supersymmetric graviton operator
$g_{i}^{(\hat{h}_{i \mu_i})} (x, \th)$ defined in (\ref{4-4}).
In terms of (\ref{4-19}) and (\ref{4-20}), the new graviton operator
$g_{\mu_i}^{(\hat{h}_{\mu_i \la_i})} (x, \th , x^\prime, \th^\prime )$
can be expressed as
\beqar
    g_{\mu_i}^{(\hat{h}_{\mu_i \la_i})} (x, \th , x^\prime, \th^\prime )
    \!\!\! &=& \!\!\!\!
    \left. \int d \mu ( p_{\mu_i} ) d \mu ( p_{\la_i} )
    ~ g_{\mu_i}^{(\hat{h}_{\mu_i \la_i})} ( \xi_{\mu_i} , \eta_{\la_i} )
    ~  e^{ i x \cdot p_{\mu_i} }e^{ i x^\prime \cdot p_{\la_i} }
    \right|_{ \xi_{\mu_i}^{\al} = \th_{A}^{\al} u_{\mu_i}^{A}, \,
    \eta_{\la_i}^{\bt} = {\th^\prime}_{B}^{\bt} u_{\la_i}^{B}  }
    \label{4-23}
    \\
    g_{\mu_i}^{(\hat{h}_{\mu_i \la_i})} ( \xi_{\mu_i} , \eta_{\la_i} )
    \!\!\! &=& \!\!\!
    T^{\la_i} \, g_{\mu_i \la_i}^{(\hat{h}_{\mu_i \la_i})} ( \xi_{\mu_i} , \eta_{\la_i} )
    \, = \,
    T^{\la_i} \, e_{\mu_i}^{(\hat{h}_{\mu_i}) a} ( \xi_{\mu_i} ) \,
    e_{\la_i}^{(\hat{h}_{\la_i}) a} ( \eta_{\la_i} )
    \label{4-24}
\eeqar
where $g_{\mu_i}^{(\hat{h}_{\mu_i \la_i})} ( \xi_{\mu_i} , \eta_{\la_i} )$
in the second equation is an analog of $g_{i}^{(\hat{h}_{i \mu_i})} ( \xi_{i} )$
in (\ref{4-5}) with two types of Grassmann variables (\ref{4-17}) and (\ref{4-18}).
The supersymmetric frame fields $e_{\mu_i}^{(\hat{h}_{\mu_i}) a} ( \xi_{\mu_i} )$
are now define by
\beqar
    \nonumber
    e_{\mu_i}^{(+)a} (\xi_{\mu_i}) &=& e_{\mu_i}^{(+) a} \\ \nonumber
    e_{\mu_i}^{\left( + \hf \right) a} (\xi_{\mu_i}) &=& \xi_{{\mu_i}}^{\al_1}
    \, e_{\mu_i \, \al_1}^{ \left( + \hf \right) a} \\
    e_{\mu_i}^{(0)a} (\xi_{\mu_i}) &=& \hf \xi_{\mu_i}^{\al_1}
    \xi_{\mu_i}^{\al_2} \, e_{\mu_i \, \al_1 \al_2}^{(0)a}
    \label{4-25}
    \\ \nonumber
    e_{\mu_i}^{\left(- \hf \right)a} (\xi_{\mu_i}) &=& \frac{1}{3!}
    \xi_{\mu_i}^{\al_1} \xi_{\mu_i}^{\al_2} \xi_{\mu_i}^{\al_3}
    \ep_{\al_1 \al_2 \al_3 \al_4} \, {e_{\mu_i}^{ \al_4}}^{ \left( - \hf \right) a}
    \\ \nonumber
    e_{\mu_i}^{(-)a} (\xi_{\mu_i}) &=& \xi_{\mu_i}^{1}
    \xi_{\mu_i}^{2} \xi_{\mu_i}^{3} \xi_{\mu_i}^{4} \, e_{\mu_i}^{(-)a}
\eeqar
where each of $\al_1 ,  \al_2 , \al_3 , \cdots$ takes a value of 1, 2, 3 or 4.
Similarly, the other set of the supersymmetric frame fields are defined by
\beqar
    \nonumber
    e_{\la_i}^{(+)a} (\eta_{\la_i}) &=& e_{\la_i}^{(+) a} \\ \nonumber
    e_{\la_i}^{\left( + \hf \right) a} (\eta_{\la_i}) &=& \eta_{\la_i}^{\bt_1}
    \, e_{\la_i \, \bt_1}^{ \left( + \hf \right) a} \\
    e_{\la_i}^{(0)a} (\la_i) &=& \hf \eta_{\la_i}^{\bt_1} \eta_{\la_i}^{\bt_2} \,
    e_{\la_i \, \bt_1 \bt_2}^{(0)a}
    \label{4-26}
    \\ \nonumber
    e_{\la_i}^{\left(- \hf \right)a} (\eta_{\la_i}) &=& \frac{1}{3!} \eta_{\la_i}^{\bt_1}
    \eta_{\la_i}^{\bt_2} \eta_{\la_i}^{\bt_3}
    \ep_{\bt_1 \bt_2 \bt_3 \bt_4} \, {e_{\la_i}^{ \bt_4}}^{ \left( - \hf \right) a}
    \\ \nonumber
    e_{\la_i}^{(-)a} (\eta_{\la_i}) &=& \eta_{\la_i}^{1} \eta_{\la_i}^{2}
    \eta_{\la_i}^{3} \eta_{\la_i}^{4} \, e_{\la_i}^{(-)a}
\eeqar
where each of $\bt_1 ,  \bt_2 , \bt_3 , \cdots$ takes a value of 1, 2, 3 or 4.

{\it
A supersymmetric gravitational holonomy operator
$\Theta_{R , \ga}^{(H)} ( u , \bu ; x , \th , x^\prime , \th^\prime )$
is then  defined by substitution of
$g_{i}^{(\hat{h}_{i \mu_i})} (x, \th , x^\prime , \th^\prime )$
into $g_{i}^{(h_{i \mu_i} )}$ in (\ref{2-2}).
}

The operator $\Theta_{R , \ga}^{(H)} ( u , \bu ; x , \th , x^\prime , \th^\prime )$
can also be written as a supersymmetrization of the expression in (\ref{3-6}), {\it i.e.},
\beqar
    \Theta_{R, \ga}^{(H)} ( u, \bu  ; x , \th , x^\prime , \th^\prime )
    & = &
    \sum_{\{ \si , \tau \}} \,
    \Theta_{R_{\mu \la}, \ga_{ \si | \tau }}^{(H)} ( u, \bu  ; x , \th , x^\prime , \th^\prime )
    \nonumber \\
    & = &
    \sum_{\{ \si , \tau \}} \,
    \left\langle \Theta_{R_\mu , \ga  }^{(E)} ( u ; x , \th ) \cdot
    \Theta_{R_\la , \ga  }^{(E)} ( u ; x^\prime , \th^\prime )
    \right \rangle_{\ga = \ga_{\si | \tau }}
    \label{4-27}
\eeqar
where $\Theta_{R_\mu , \ga_{\si | \tau }  }^{(E)} ( u ; x , \th ) $ and
$\Theta_{R_\la , \ga_{\si | \tau } }^{(E)} ( u ; x^\prime , \th^\prime )$ are defined as follows.
{\it
\begin{enumerate}
  \item $\Theta_{R_\mu , \ga_{\si | \tau }  }^{(E)} ( u ; x , \th )$ is obtained by
  substitution of $e_{\mu_i}^{( \hat{h}_{\mu_i} ) } ( x , \th )$ into
  $e_{\mu_i}^{(h_{\mu_i})}$ in the definition of
  $\Theta_{R_\mu , \ga_{\si | \tau }  }^{(E)} ( u )$ given by (\ref{3-7}).
  An explicit expansion form of (\ref{3-7}) can be obtained from (\ref{3-4}) and (\ref{2-30}).
  Notice that the frame-field operator $e_{\mu_i}^{(h_{\mu_i})}$ enters in
  the comprehensive frame field $E$ as defined in (\ref{2-26}) and (\ref{2-27}).
  \item $\Theta_{R_\la , \ga  }^{(E)} ( u ; x^\prime , \th^\prime )$ is obtained by
  substitution of $e_{\la_i}^{( \hat{h}_{\la_i} ) } ( x^\prime , \th^\prime )$ into
  $e_{\mu_i}^{(h_{\mu_i})}$ in the definition of
  $\Theta_{R_\la , \ga_{\si | \tau } }^{(E)} ( u )$ given by (\ref{3-8}).
\end{enumerate}
}

Using the squared holonomy operator $\Theta_{R_{\mu \la}, \ga_{ \si | \tau }}^{(H)}
( u, \bu  ; x , \th , x^\prime , \th^\prime )$
in (\ref{4-27}), we can define an S-matrix functional for a subset of
the MHV graviton amplitudes as
\beq
    \F_{MHV} \left[  e_{\mu_i}^{(h_{\mu_i})} \cdot e_{\la_i}^{(h_{ \la_i})} \right]
    \, = \, \exp \left[ \frac{i}{8 \pi G_N} \int d^4 x \, d^{8} \th \, d^{8} \th^\prime
    ~
    \Theta_{R_{\mu \la}, \ga_{ \si | \tau }}^{(H)} ( u, \bu  ; x , \th , x^\prime , \th^\prime )
    \right]
    \label{4-28}
\eeq
where $ e_{\mu_i}^{(h_{\mu_i})} \cdot e_{\la_i}^{(h_{ \la_i})}$ denotes a source
function that is associated with the composite operator
$g_{\mu_i \la_i}^{( h_{\mu_i \la_i} )} = e_{\mu_i}^{(h_{\mu_i})} \cdot e_{\la_i}^{(h_{ \la_i})}$.
For simplicity, we here express a product of the frame fields on the tangent space
by a dot product rather than using the tangent-space index $a$ $(=0,1,2,3)$.

As reviewed in the appendix, the graviton amplitudes of arbitrary helicity configuration
can generally be expressed in the form of (\ref{A-10}).
Applying this expression to the MHV graviton amplitudes,
we can easily check that
the amplitudes $\widehat{M}_{MHV}^{(s_{--} t_{--} )} (u, \bu)$ can be generated by
$\F_{MHV} \left[  e_{\mu_i}^{(h_{\mu_i})} \cdot e_{\la_i}^{(h_{ \la_i})} \right]$ as follows.
\beqar
    &&
    \left.
    \sum_{ \{ \si , \tau \}}
    \left[ \bigotimes_{( s_{-} t_{-} )} \frac{\del}{\del e_{\mu_i}^{(h_{\mu_i} )}( x )}
    \right] \cdot
    \left[ \bigotimes_{( s_{-} t_{-} )} \frac{\del}{\del e_{\la_i}^{(h_{\la_i} )} (x^\prime ) }
    \right]
    ~ \F_{MHV} \left[ e_{\mu_i}^{(h_{\mu_i})} \cdot e_{\la_i}^{(h_{ \la_i})} \right]
     \right|_{ e_{\mu_i}^{(h_{\mu_i})} (x) = e_{\la_i}^{(h_{ \la_i})}(x^\prime ) = 0 }
    \nonumber \\
    & = &
    i ( 8 \pi G_N )^{\frac{n}{2}-1} \, \widehat{M}_{MHV}^{(s_{--} t_{--})} (u , \bu)
    \label{4-29}
\eeqar
where the (direct) product sums of the functional derivatives are defined by
\beqar
    \!\!\!\!\!
    \bigotimes_{( s_{-} , t_{-} )} \frac{\del}{\del e_{\mu_i}^{(h_{\mu_i} )}( x )}
    \!\!\! & \equiv & \!\!\!
    \frac{\del}{\del e_{\mu_1}^{(+)}( x )} \otimes
    \cdots \otimes \frac{\del}{\del e_{ \mu_s}^{(-) }( x )} \otimes
    \cdots \otimes \frac{\del}{\del e_{ \mu_t}^{(-) }( x )} \otimes
    \cdots \otimes \frac{\del}{\del e_{\mu_n}^{(+)}( x )}  \, ,
    \label{4-30} \\
    \!\!\!\!\!
    \bigotimes_{( s_{-} , t_{-} )} \frac{\del}{\del e_{\la_i}^{(h_{\la_i} )}(x^\prime )}
    \!\!\! & \equiv & \!\!\!
    \frac{\del}{\del e_{\la_1}^{(+)}(x^\prime )} \otimes
    \cdots \otimes \frac{\del}{\del e_{ \la_s}^{(-) }(x^\prime )} \otimes
    \cdots \otimes \frac{\del}{\del e_{ \la_t}^{(-) }(x^\prime )} \otimes
    \cdots \otimes \frac{\del}{\del e_{ \la_n}^{(+) }(x^\prime )} \, .
    \label{4-31}
\eeqar
The expression (\ref{4-29}) is a homogeneous-sum version of the expression (\ref{4-12}).
Notice that the Grassmann integrals over $\th$ and $\th^\prime$
pick up only the MHV-type helicity configuration since the
integrals vanish unless we have the following factors:
\beqar
    \left. \int d^8 \th  \, \xi_{\mu_s}^{1}\xi_{\mu_s}^{2}\xi_{\mu_s}^{3}\xi_{\mu_s}^{4}
    \, \xi_{\mu_t}^{1}\xi_{\mu_t}^{2}\xi_{\mu_t}^{3}\xi_{\mu_t}^{4}
    \right|_{\xi_{\mu_i}^{\al} = \th_{A}^{\al} u_{\mu_i}^{A} }
    &=&  (u_{\mu_s} u_{\mu_t} )^4 \, ,
    \label{4-32} \\
    \left. \int d^8 \th^\prime  \, \eta_{\la_s}^{1}\eta_{\la_s}^{2}\eta_{\la_s}^{3}
    \eta_{\la_s}^{4} \, \eta_{\la_t}^{1}\eta_{\la_t}^{2}\eta_{\la_t}^{3}\eta_{\la_t}^{4}
    \right|_{\eta_{\la_i}^{\bt} = {\th^\prime}_{B}^{\bt} u_{\la_i}^{B} }
    &=&  (u_{\la_s} u_{\la_t} )^4 \, .
    \label{4-33}
\eeqar
In the homogeneous-sum case, as shown in (\ref{3-1}),
the gravitational operator is denoted by $T^{\la_i} g_{\mu_i \la_i}^{( h_{\mu_i \la_i} )}
= g_{\mu_i}^{( h_{\mu_i \la_i} )}$. Thus the helicity of the $\mu_i$-th
graviton is labeled by $ h_{\mu_i \la_i} $.
In a practical calculation of (\ref{4-29}), we first
set $( \si_2 , \si_3 , \cdots , \si_r )= ( 2,3,\cdots , r)$ and
$( \tau_{r+1} , \tau_{r+2} , \cdots ,\tau_{n-2} ) = ( r+1 ,r+2 , \cdots , n-2)$,
or equivalently $( \mu_1 , \mu_2 , \cdots , \mu_n ) = ( 1, 2, \cdots , n )$
and $( \la_1 , \la_2 , \cdots , \la_n ) = ( 2,3, \cdots, r , 1, n-1 , r+1 , r+2 ,
\cdots , n-2, n )$, and then take the double permutation of $\P ( \si , \tau )$ before
carrying out the homogeneous sum where the helicity information is synchronized with
the numbering indices.
Thus the choices of the functional derivatives (\ref{4-30}), (\ref{4-31}) correctly lead
to the MHV configurations of the amplitudes $\widehat{M}_{MHV}^{(s_{--} t_{--})} (u , \bu)$.

In the $x$-space representation, the MHV amplitudes can be generated as
\beqar
    &&
    \left.
    \sum_{ \{ \si , \tau \}}
    \left[ \bigotimes_{( s_{-} t_{-} )} \frac{\del}{\del e_{\mu_i}^{(h_{\mu_i} )}}
    \right] \cdot
    \left[ \bigotimes_{( s_{-} t_{-} )} \frac{\del}{\del e_{\la_i}^{(h_{\la_i} )}(x^\prime ) }
    \right]
     \F_{MHV} \left[ e_{\mu_i}^{(h_{\mu_i})} \cdot e_{\la_i}^{(h_{ \la_i})}
    \right]
     \right|_{ e_{\mu_i}^{(h_{\mu_i})} = e_{\la_i}^{(h_{ \la_i})} (x^\prime ) = 0 }
    \nonumber \\
    & = &
    (-1)^{n+1} \, \M_{MHV}^{(s_{--} t_{--})} (x) \, .
    \label{4-34}
\eeqar
As in the case of (\ref{4-16}), the energy-conservation delta function
naturally arises from the functional derivatives with respect to
$e_{\mu_i}^{(h_{\mu_i} )}$'s.
Notice that in this representation the momenta of gravitons
are equivalent to those of the frame fields;
we do not have to make the latter momenta be half of the former
as usually prescribed for the momenta of closed and open strings
in superstring theory.

Obtaining the expression (\ref{4-29}) or (\ref{4-34}) is the main objective of the present paper.
It shows that an S-matrix functional for the MHV graviton amplitudes can also be
described in terms of the supersymmetric gravitational holonomy operator,
$\Theta_{R_{\mu \la}, \ga_{ \si | \tau }}^{(H)} ( u, \bu  ; x , \th , x^\prime , \th^\prime )$
defined in (\ref{4-27}), by use of the homogeneous sum.
This gives a concrete realization of the rough idea that gravity can be considered as a square
of gauge theory at the level of construction of the holonomy operators in twistor space.
For the completion of our analysis, we shall consider a generalization to
the non-MHV amplitudes in what follows..

\noindent
\underline{The non-MHV amplitudes}
%\subsection{Non-MHV amplitudes}

Generalization of the above analysis to the non-MHV amplitudes can
be carried out straightforwardly by use of the so-called
Cachazo-Svrcek-Witten (CSW) rules \cite{Cachazo:2004kj}.
The rules are summarized by the expressions (\ref{A-3}) and (\ref{A-4})
in the appendix.
In the language of functional integrals, these rules can
be realized succinctly by use of the following S-matrix functional \cite{Abe:2009kq}:
\beqar
    \F \left[ g_{i \mu_i}^{(h_{i \mu_i})} \right]
    &= &  \widehat{W}^{(H)} \, \F_{MHV} \left[ g_{i \mu_i}^{(h_{i \mu_i})} \right]
    \label{4-35}\\
    \widehat{W}^{(H)} &=& \exp \left[
    \int d^4 x d^4 y ~ \frac{\del_{kl}}{q^2} ~
    \frac{\del}{\del g_{k \mu_k}^{(++)}(x)} \otimes
    \frac{\del}{\del g_{l \mu_l}^{(--)}(y)}
      \right]
    \label{4-36}
\eeqar
where $q$ in (\ref{4-36}) is a momentum transferred between the vertices at $x$ and $y$.
This momentum transfer plays the same role as $q_{ij}$ in (\ref{A-4}) for the next-to-MHV amplitudes.
(The contraction operator (\ref{4-36}) that realizes the CSW rules is first
introduced in \cite{Abe:2004ep} for gluon amplitudes.)
The general S-matrix functional (\ref{4-35}) is defined
in terms of the MHV S-matrix functional (\ref{4-8}) for the split-sum case.

Using (\ref{4-35}), we can
generate tree-level graviton amplitudes in general as
\beqar
    &&
    \nonumber
    \left. \frac{\del}{\del g_{1 \mu_1}^{(h_{1 \mu_1 } )} (x_1) } \otimes
    \frac{\del}{\del g_{2 \mu_2}^{(h_{2 \mu_2}) } (x_2)} \otimes
    \cdots \otimes \frac{\del}{\del g_{n \mu_n}^{(h_{n \mu_n}) } (x_n)}
    ~ \F \left[  g_{i \mu_i}^{(h_{i \mu_i})}  \right]
    \right|_{g_{i \mu_i}^{(h_{i \mu_i})} (x)=0} \\
    &=&
    i ( 8 \pi G_N )^{\frac{n}{2}-1}
    \widehat{M}^{(1_{h_{1 \mu_1}} 2_{h_{2 \mu_2}} \cdots n_{h_{n \mu_n}})} (u , \bu )
    \label{4-37}
\eeqar
where the helicity $h_{i \mu_i}$ $(i = 1, 2, \cdots , n)$ takes a value of $(++ , --)$.
Other configurations, such as $(+-, -+)$, are ruled out
due to the Grassmann integral in (\ref{4-13}).

Notice that the particular assignment for the
index $\al$ in (\ref{4-3}) is crucial to extract the helicities of $(++,--)$.
Without such an assignment, particles with $(+-, -+)$ helicities would emerge.
The operators $g_{i \mu_i}^{(+-)}$
and $g_{i \mu_i}^{(-+)}$ are to represent stable and electrically neutral
particles without mass or spin which we may regard
as candidates for the origin of dark matter.
Although this is nothing but an intuitive speculation,
observational evidence of dark matter and dark energy suggests
that there might be operators like
$g_{i \mu_i}^{(+-)}$ and $g_{i \mu_i}^{(-+)}$
to be incorporated in a full gravitational theory.
In the present formalism, this can be carried out by relaxing
the assignment (\ref{4-3}) and using instead the assignments
of (\ref{4-17}), (\ref{4-18}); this lead to a gravitational theory as a square
of an $\N =4$ supersymmetric gauge theory for frame fields such that
the gravitational theory includes operators involving
$g_{i \mu_i}^{(+-)}$ and $g_{i \mu_i}^{(-+)}$.

At the level of the construction of
the MHV S-matrix functional, such a treatment
can be made by using
$\F_{MHV} \left[  e_{\mu_i}^{(h_{\mu_i})} \cdot e_{\la_i}^{(h_{ \la_i})} \right]$
in (\ref{4-28}) rather than
$\F_{MHV} \left[ g_{i \mu_i}^{(h_{i \mu_i})} \right]$ in (\ref{4-8}).
Using the former, in comparison to the forms in (\ref{4-35})-(\ref{4-37}),
we can define an alternative expression for the non-MHV S-matrix functional:
\beqar
    \F \left[  e_{\mu_i}^{(h_{\mu_i})} \cdot e_{\la_i}^{(h_{ \la_i})} \right]
    &= &  \widehat{W}^{(EE)} \, \F_{MHV}
    \left[  e_{\mu_i}^{(h_{\mu_i})} \cdot e_{\la_i}^{(h_{ \la_i})} \right]
    \label{4-38}\\
    \widehat{W}^{(EE)} &=& \exp \left[
    \int d^4 x d^4 y ~ \frac{\del_{kl}}{q^2} \,
    \frac{\del}{\del e_{\mu_k}^{(+)}(x)} \otimes
    \frac{\del}{\del e_{\mu_l}^{(-)}(y)}
    \right]
    \nonumber \\
    &&
    \cdot \,
    \exp \left[
    \int d^4 x^\prime d^4 y^\prime ~ \frac{ \del_{k^\prime l^\prime}}{{q^\prime}^2} \,
    \frac{\del}{\del e_{\la_{k^\prime}}^{(+)} ( x^\prime ) } \otimes
    \frac{\del}{\del e_{\la_{l^\prime}}^{(-)} ( y^\prime ) }
    \right]
    \label{4-39}
\eeqar
where $q$ ($q^\prime$) denotes a momentum transfer
between the vertices at $x$ ($x^\prime$) and $y$ ($y^\prime$).
Notice that, in the homogeneous-sum case, graviton amplitudes are factorized
by the MHV vertices for frame fields while, in the split-sum case, they are factorized
by the graviton MHV vertices.
This explains why we have two $q$'s in (\ref{4-39})
while there is a single $q$ in (\ref{4-36}).

Using the new S-matrix functional (\ref{4-38}), we can also
generate the non-MHV graviton amplitudes as
\beqar
    &&
    \left.
    \sum_{ \{ \si , \tau \}}
    \left[ \bigotimes \frac{\del}{\del e_{\mu_i}^{(h_{\mu_i} )}( x )}
    \right] \cdot
    \left[ \bigotimes \frac{\del}{\del e_{\la_i}^{(h_{\la_i} )} ( x^\prime )}
    \right]
    ~ \F \left[ e_{\mu_i}^{(h_{\mu_i})} \cdot e_{\la_i}^{(h_{ \la_i})} \right]
     \right|_{ e_{\mu_i}^{(h_{\mu_i})} (x) = e_{\la_i}^{(h_{ \la_i})} ( x^\prime )= 0 }
    \nonumber \\
    & = &
    i ( 8 \pi G_N )^{\frac{n}{2}-1}
    \widehat{M}^{(1_{h_{\mu_1 \la_1}} 2_{h_{\mu_2 \la_2}} \cdots n_{h_{\mu_n \la_n}})} (u , \bu )
    \label{4-40}
\eeqar
where the sets of functional derivatives are now defined by
\beqar
    \!\!
    \bigotimes \frac{\del}{\del e_{\mu_i}^{(h_{\mu_i} )}( x )}
    \!\! & \equiv & \!\!
    \frac{\del}{\del e_{\mu_1}^{( h_{\mu_1})}( x_1 )} \otimes
    \frac{\del}{\del e_{\mu_2}^{( h_{\mu_2})}( x_2 )} \otimes
    \cdots \otimes \frac{\del}{\del e_{\mu_n}^{( h_{\mu_n})}( x_n )} \, ,
    \label{4-41} \\
    \!\!
    \bigotimes \frac{\del}{\del e_{\la_i}^{(h_{\la_i} )}( x^\prime )}
    \!\! & \equiv & \!\!
    \frac{\del}{\del e_{\mu_1}^{( h_{\la_1})} ( x^{\prime}_{1} )} \otimes
    \frac{\del}{\del e_{\mu_2}^{( h_{\la_2})} ( x^{\prime}_{2} )} \otimes
    \cdots \otimes \frac{\del}{\del e_{\la_n}^{( h_{\la_n})}( x^{\prime}_{n} )} \, .
    \label{4-42}
\eeqar
The expression (\ref{4-40}) confirms that the non-MHV S-matrix functional
can indeed be obtained in terms of the holonomy operator
$\Theta_{R_{\mu \la}, \ga_{ \si | \tau }}^{(H)} ( u, \bu  ; x , \th , x^\prime , \th^\prime )$
defined in (\ref{4-27}).
Notice that the helicities $h_{\mu_i \la_i} = h_{\mu_i} h_{\la_i}$ ($i= 1,2, \cdots , n$)
in (\ref{4-40}) can take any combinations including $(+- , -+)$.
Since graviton operators are defined by
$g_{\mu_i \la_i}^{( h_{\mu_i \la_i} )} = e_{\mu_i}^{(h_{\mu_i})} \cdot e_{\la_i}^{(h_{ \la_i})}$,
this means that the above formulation suggests the
existence of particles labeled by
$g_{\mu_i \la_i}^{(+-)}$ and $g_{\mu_i \la_i}^{(-+)}$.

%\section{Discussions}
\section{Concluding remarks}

In the present paper, we further consider a gravitational holonomy operator
$\Theta_{R, \ga}^{(H)} (u , \bu )$ that has been developed in \cite{Abe:2009kq}.
The holonomy operator is defined in twistor space, with $u$, $\bu$ denoting
spinor momenta of gravitons defined in a $\cp^1$-fiber of the twistor space.
In section 2, we first review the construction of $\Theta_{R, \ga}^{(H)} (u , \bu )$ and
how it can be interpreted as a holonomy operator of gauge fields with
a certain combinatoric Chan-Paton factor.
The structure of the Chan-Paton factor, explicitly given in
(\ref{2-11})-(\ref{2-14}), is the same as the structure of a Chan-Paton factor
in graviton amplitudes that has been obtained by Bern et al. in \cite{Bern:1998sv}.
As shown in \cite{Abe:2009kq}, this relation is utilized to obtain an S-matrix functional for
graviton amplitudes in terms of a supersymmetric version of
the gravitational holonomy operator, $\Theta_{R, \ga}^{(H)} (u , \bu; x, \th )$, with $\N =8$
extended supersymmetry.
Here $x$ denotes a spacetime coordinate that emerges from the twistor space
and $\th^\al$ ($\al = 1,2, \cdots 8$) denotes Grassmann variables that compose
$\N =8$ chiral superspace.
The construction of such an S-matrix functional is also reviewed in section 4.

We present the main results of this paper in section 3.
There we give an alternative expression for a gravitational holonomy operator
such that it can be interpreted as a square of an ${\cal N}=4$ holonomy operator
for frame fields.
The expression is motivated by the previous work \cite{Abe:2005se}
and is obtained by use of what we call the homogeneous sum.
This sum is taken by certain shuffles over ordered numbering indices.
An explicit form of $\Theta_{R, \ga}^{(H)} (u , \bu )$ with
such a sum is shown in (\ref{3-6}).
Supersymmetrization of this expression,
$\Theta_{R, \ga}^{(H)} (u , \bu; x, \th , x^\prime , \th^\prime )$,
is considered in section 4 and is explicitly shown in (\ref{4-27}).
In section 4, we also show that this squared expression can also be used to
define an S-matrix functional for the general non-MHV graviton amplitudes.

The homogeneous sum that appears in the new expression is
equivalent to a sum over what is called $(k,l)$-shuffles ($k, l \in {\bf N}$)
in mathematics. Such a sum appears, for example, in
(a) the definition of Laplace expansion formula for
determinants in terms of the so-called Pl\"{u}cker coordinates,
(b) the definition of a wedge product of a differential $k$-form
and a differential $l$-form, and (c) the definition of a product of iterated integrals
defined by a set of differential one-forms, say, $\om_1 , \om_2 , \cdots , \om_{k+l}$.
We have seen an explicit definition for the case of (c) in (\ref{3-9}).
In (\ref{3-9})-(\ref{3-12}), we then argue that the gravitational holonomy operator can be
interpreted as a square of a same theory which is represented by
a frame-field holonomy operator.
Regarding the case of (a), it suggests that $\Theta_{R, \ga}^{(H)} (u , \bu )$
may be interpreted as some determinant.
The holonomy operator is related to a Wess-Zumino-Witten (WZW) action,
or more precisely to the current correlator of a WZW model.
The WZW action, on the other hand, is closely related to a chiral Dirac determinant.
(For the relation between the WZW action and the gluon amplitudes in this context,
see \cite{Abe:2004ep}.)
It is then natural to interpret $\Theta_{R, \ga}^{(H)} (u , \bu )$
as a chiral Dirac determinant suitably defined in twistor space.
Details of this relation are currently under study.

Lastly, we would like to discuss that the squared expression
we obtain in this paper is theoretically more natural
than the previously known expression.
In the holonomy formalism of gravity, we need to take essentially two sums.
One is a sum over all possible metrics that guarantees
general covariance of the theory, and the other is
a sum over permutations of the numbering indices, or a braid trace,
that guarantees diffeomorphism invariance of the theory.
In the squared expression, the former sum is realized by
the homogeneous sum and the latter is realized by double-permutation
terms, denoted as $\P (\si | \tau ) + \P (\si | \tau )$, in (\ref{3-1}).
The double permutation can also be written as a double braid-trace in (\ref{3-5}).
On the other hand, in the original expression (\ref{2-32}),
the sum over metrics are realized by the split sum
$\sum_{\si \in \S_{r-1}} \sum_{\tau \in \S_{m-r-1}}$ and
the braid trace is realized by $\P ( 2 3 \cdots m-2 )$.
Thus, for either case, we can make
physically clear interpretations to the sums that appear
in the expressions of $\Theta_{R, \ga}^{(H)} (u , \bu )$.
These interpretations are summarized in Table \ref{table01}.

Our preference for the squared expression
arises upon supersymmetrization of $\Theta_{R, \ga}^{(H)} (u , \bu )$.
In the split-sum case, the supersymmetric operator is defined by
$\Theta_{R, \ga}^{(H)} (u , \bu; x, \th )$,
with Grassmann variables $\th^\al$ ($\al = 1,2, \cdots 8$)
split into two parts, $\al_1 = 1,2,3,4$ and $\al_2 = 5,6,7,8$,
as shown in (\ref{4-3}).
Although this setting leads to the correct graviton amplitudes,
there are no {\it a priori} reasons to choose this particular splitting.
In this sense, it is an artificial setting.
This problem does not occur in the homogeneous-sum case where
the supersymmetric operator is defined by
$\Theta_{R, \ga}^{(H)} (u , \bu; x, \th , x^\prime , \th^\prime)$,
with Grassmann variables,
$\th^\al$ ($\al = 1,2,3,4$) and ${\th^\prime}^{\bt}$ ($\bt = 1,2,3,4$),
as shown in (\ref{4-17}) and (\ref{4-18}).
In this case, there are no restrictions on the indices $\al$ and $\bt$.
As a consequence, this theory contains particles of
helicity configuration $(+- , -+ )$
in addition to the pure-gravity helicity configuration $(++ , --)$.
The extra particles are massless spin-zero particles with no electric charges.
These particles are also expected to be stable as the ordinary gravitons.
Thus we can naturally interpret these as candidates for the origin of dark matter.
Research on this speculative idea will be reported in a future paper.

There is another theoretical reason to prefer the squared theory to the original one.
In the holonomy formalism, the physical variables are given by a
set of spinor momenta defined on a $\cp^1$ fiber of twistor space $\cp^3$.
An underlining space of interest is thus the twistor space
without which we would not construct physical operators in four-dimensional spacetime.
In other words, in analogy with the language of a WZW model, we can
define and identify a target space of the holonomy operator by the twistor space.
In terms of this terminology, the target space of the supersymmetric holonomy operator
$\Theta_{R, \ga}^{(H)} (u , \bu; x, \th )$ is given by $\cp^{3|8}$, while that of
$\Theta_{R, \ga}^{(H)} (u , \bu; x, \th , x^\prime , \th^\prime)$
is given by $\cp^{3|4} \times \cp^{3|4}$.
Notice that $\cp^{3|4}$ is a super Calabi-Yau manifold but $\cp^{3|8}$ is not.
This means that one can construct a superstring theory, which of course contains
quantum gravity, on $\cp^{3|4} \times \cp^{3|4}$ but not on $\cp^{3|8}$.
Thus, from this perspective as well, it is natural to favor the squared theory.

%%%%%%%%%%%%%%%%%%%%%%%%%%%%%%%%%%%%
\vskip .3in
\noindent
{\bf Acknowledgments} \vskip .06in
\noindent
The author would like to thank Professor V.P. Nair for comments on the previous
work \cite{Abe:2005se} which have been useful for the present paper.

%%%%%%%%%%%%%%%%%%%%%%%%%%%%%%%%%%%%

\appendix
\section{Graviton amplitudes}
%\section{Appendix: Graviton amplitudes}

In this appendix, we review some expressions of
graviton amplitudes in relation to those of gluon counterparts.
Most of the following results are obtained in \cite{Abe:2009kn,Abe:2009kq}.
We here simply give those expressions that are of direct relevance to the present paper.

\noindent
\underline{Gluon amplitudes, MHV amplitudes and the CSW rules}

We first consider the gluon amplitudes.
In the spinor-momenta formalism, the simplest way of
describing the gluon amplitudes is to factorize the amplitudes
in terms of the maximally helicity violating (MHV) amplitudes.
The MHV amplitudes are the scattering amplitudes of $(n-2)$
positive-helicity gluons and $2$ negative-helicity gluons or the other way around.
In a momentum-space representation, the MHV tree amplitudes of gluons
are expressed as
\beqar
    \nonumber
    \A_{MHV}^{(1_+ 2_+ \cdots r_{-} \cdots s_{-} \cdots n_+ )} (u, \bu)
    & \equiv &
    \A_{MHV}^{(r_{-} s_{-})} (u, \bu) \\
    & = & i g^{n-2}
    \, (2 \pi)^4 \del^{(4)} \left( \sum_{i=1}^{n} p_i \right) \,
    \widehat{A}_{MHV}^{(r_{-} s_{-})} (u)
    \label{A-1}
    \\
    \widehat{A}_{MHV}^{(r_{-} s_{-})} (u) \! &=&
    \!\! \sum_{\si \in \S_{n-1}} \!\!
    \Tr (t^{c_1} t^{c_{\si_2}} t^{c_{\si_3}} \cdots t^{c_{\si_n}}) \,
    \frac{ (u_r u_s )^4}{ (u_1 u_{\si_2})(u_{\si_2} u_{\si_3})
    \cdots (u_{\si_n} u_1)}
    \label{A-2}
\eeqar
where the elements $r$ and $s$ denote the numbering indices of the negative-helicity gluons,
$g$ represents the Yang-Mills coupling constant, and $t_{c_i}$'s are
the Chan-Paton factors of gluons.
$u_i$ denotes the two-component spinor momentum of the $i$-th gluon ($i=1,2, \cdots, n$).
In terms of $u_i^A$ ($A = 1,2$) and its complex conjugate $\bu_i^\Ad$
($\Ad = 1,2$), the four-dimensional gluon momentum $p_{i}^{A \Ad}$ is parametrized by
\beq
    p_{i}^{A \Ad} \, = \, u_i^A \bu_i^\Ad \, .
    \label{A-2a}
\eeq
This parametrization is explicitly shown in (\ref{2-17}) and (\ref{2-18}).
Of particular interest in the spinor-momenta formalism is that the MHV gluon amplitudes
$\widehat{A}_{MHV}^{(r_{-} s_{-})} (u) $ is purely holomorphic in terms of the
spinor momentum $u_i$. For the MHV graviton amplitudes, however, it no longer holds
since the Chan-Paton factors of gravitons in the spinor-momenta formalism
are composed of a set of four-dimensional graviton momenta analogous to (\ref{A-2a}).
An explicit form of the MHV graviton amplitudes is given in (\ref{4-10}).

The non-MHV gluon amplitudes, or the general gluon amplitudes, can be expressed in terms of
the MHV amplitudes $\widehat{A}_{MHV}^{(r_{-} s_{-})} (u)$.
Prescription for such expressions is called the Cachazo-Svrcek-Witten (CSW) rules
\cite{Cachazo:2004kj}.
For the next-to-MHV (NMHV) amplitudes, which contain 3 negative-helicity
gluons and $(n-3)$ positive-helicity gluons, the CSW rules can be expressed as
\beq
    \widehat{A}^{(r_- s_- t_-)}_{NMHV} (u) = \sum_{(i,j)}
    \widehat{A}^{(i_+ \cdots r_- \cdots s_- \cdots j_+ k_+)}_{MHV} (u)
    \, \frac{\del_{kl}}{q_{ij}^2} \,
    \widehat{A}^{(l_- \, (j+1)_+ \cdots t_- \cdots (i-1)_+)}_{MHV} (u)
    \label{A-3}
\eeq
where the sum is taken over all possible choices for $(i, j)$ that
satisfy the ordering $i < r < s < j < t$ (mod $n$).
The numbering indices for the negative-helicity gluons are now given by
$r$, $s$ and $t$.
The momentum transfer $q_{ij}$ between the two MHV vertices can be
expressed by a set of the gluon four-momenta:
\beq
    q_{ij} = p_i + p_{i+1 } + \cdots + p_{r} + \cdots + p_{s} + \cdots + p_{j} \, .
    \label{A-4}
\eeq
The non-MHV amplitudes are then obtained by iterative use of
the relation (\ref{A-3}).
Thus, in principle, we can express the general gluon amplitudes as
\beqar
    \A^{(1_{h_1} 2_{h_2} \cdots n_{h_n})} (u, \bu)
    & = & i g^{n-2}
    \, (2 \pi)^4 \del^{(4)} \left( \sum_{i=1}^{n} p_i \right) \,
    \widehat{A}^{(1_{h_1} 2_{h_2} \cdots n_{h_n})} (u)
    \label{A-5}
    \\
    \widehat{A}^{(1_{h_1} 2_{h_2} \cdots n_{h_n})} (u) &=&
    \sum_{\si \in \S_{n-1}}
    \Tr (t^{c_1} t^{c_{\si_2}} t^{c_{\si_3}} \cdots t^{c_{\si_n}}) ~
    \widehat{C} (1 \si_2 \si_3 \cdots \si_n)
    \label{A-6}
\eeqar
where $h_i = \pm$ denotes the helicity of the $i$-th gluon
and $\widehat{C} (1 \si_2 \si_3 \cdots \si_n)$ denotes a function of
the Lorentz-invariant scalar products $(u_i u_j)$.
The simplest form of this function is given in the case of the MHV
amplitudes:
\beq
    \widehat{C}_{MHV}^{ (r_{-} s_{-})} (1 \si_2 \si_3 \cdots \si_n) \, = \,
    \frac{ (u_r u_s )^4}{ (u_1 u_{\si_2})(u_{\si_2} u_{\si_3})
    \cdots (u_{\si_n} u_1)} \, .
    \label{A-7}
\eeq
By use of the CSW rules, we can then obtain
$\widehat{C}$'s of any helicity configurations.
Notice that $\widehat{C}_{MHV}$'s are holomorphic
in terms of the scalar products $(u_i u_j)$ but
$\widehat{C}$'s are not holomorphic in general due to
the factor of $q_{ij}$'s in (\ref{A-4}).

\noindent
\underline{Graviton amplitudes: uses of the split sum and the homogeneous sum}

In terms of such $\widehat{C}$'s, one can
express tree-level graviton amplitudes. According to \cite{Bern:1998sv},
an explicit form of the graviton amplitudes is given by
\beqar
    \nonumber
    \M^{(1_{h_{1 \mu_1}} 2_{h_{2 \mu_2}} \cdots n_{h_{n \mu_n}})} (u, \bu)
    & = &  i ( 8 \pi G_N )^{\frac{n}{2} - 1} (-1)^{n+1}
    \, (2 \pi)^4 \del^{(4)} \left( \sum_{i=1}^{n} p_i \right) \,
    \\
    &&
    ~ \times \,
    \widehat{M}^{(1_{h_{1 \mu_1}} 2_{h_{2 \mu_2}} \cdots n_{h_{n \mu_n}})}
    (u, \bu)
    \label{A-8}
    \\
    \nonumber
    \widehat{M}^{(1_{h_{1 \mu_1}} 2_{h_{2 \mu_2}} \cdots n_{h_{n \mu_n}})}
    (u, \bu)
    & = &  \sum_{\si \in \S_{r-1}} \sum_{\tau \in \S_{n-r-2}}
    \, \prod_{i=2}^{r} T^{\si_i} \, \prod_{i=r+1}^{n-1} T^{\tau_i}
    ~ \widehat{C} (1 2 \cdots n)
    \\
    \nonumber
    && \, \times \, \widehat{C} (\si_2 \si_3 \cdots \si_{r} \, 1 \, n-1 \,
    \tau_{r+1} \tau_{r+2} \cdots \tau_{n-2} \, n)
    \\
    && \, + \, \P (23 \cdots n-2)
    \label{A-9}
\eeqar
where the indices $\mu_{i}$ follow the definition (\ref{2-3}).
Also $T^{\si_i}$ and $T^{\tau_i}$ are given by (\ref{2-12}) and (\ref{2-13}),
respectively, with $m$ replaced by $n$.

The expression (\ref{A-9}) uses the split sum that we discuss in section 3.
In terms of the homogeneous sum, this can be rewritten as
\beqar
    &&
    \widehat{M}^{(1_{h_{1 \mu_1}} 2_{h_{2 \mu_2}} \cdots n_{h_{n \mu_n}})}
    (u, \bu) ~~~
    \nonumber \\
    \!\! & = & \!\!\!\!\!\!
    \sum_{\{\si,\tau \}=\{2,3, \cdots, m-2 \}}
    \!\!
    \left[
    \widehat{C} (\mu_1 \mu_2 \cdots \mu_m )
    \left( \,
    \prod_{i=1}^{m} T^{\mu_i} \,
    \widehat{C} (\la_1 \la_2 \cdots \la_m ) +  \P(\si |  \tau ) \right)  + \P(\si  |  \tau )
    \right]
    \label{A-10}
\eeqar
where the indices $\la_{i}$ follow the definition (\ref{2-4}) and
$\P (\si | \tau )$ is defined by (\ref{3-3}).
The expression (\ref{A-10}) is first obtained in \cite{Abe:2005se}.
Notice that the factor of $\prod_{i=1}^{m} T^{\mu_i}$ can be replaced by
$\prod_{i=1}^{m} T^{\la_i}$ since this product sum is determined only by
the permutations $\si$ and $\tau$.
The homogeneous sum in (\ref{A-10}) is taken over the all possible combinations for
the elements $\{ \si_2 , \si_3 , \cdots , \si_r , \tau_{r+1} ,
\tau_{r+2} , \cdots , \tau_{m-2} \}$ such that the ordering conditions
$\si_2 < \si_3 < \cdots < \si_r$ and
$\tau_{r+1} < \tau_{r+2} < \cdots < \tau_{m-2} $ are preserved.
In mathematical literature, this sum is sometimes called a sum over the $(r-1, m-2-r)$-shuffles.

%%%%%%%%%%%%%%%%%%%%%%%%%%%%%%%%%%%%%%%%%%%%%%%%%%%%%%%%%%%%%%%%

\end{document}